\documentclass[twocolumn]{aastex62}

\defcitealias{2019ApJCambioni}{Paper I}
\def\paperone{\citetalias{2019ApJCambioni}}

\defcitealias{2019ApJEmsenhuberA}{EA19a}

\usepackage{savesym}
\savesymbol{tablenum}
\usepackage[separate-uncertainty=true,multi-part-units=single]{siunitx}
\restoresymbol{SIX}{tablenum}

\usepackage{wasysym}

\DeclareSIUnit\au{au}
\DeclareSIUnit\mearth{M_\oplus}
\DeclareSIUnit\myr{Myr}

\received{2019 August 27}
\revised{2019 December 31}
\accepted{2020 January 3}
\submitjournal{ApJ}

\begin{document}

\title{Realistic On-the-fly Outcomes of Planetary Collisions II: Bringing Machine Learning to $N$-body Simulations}

\correspondingauthor{Alexandre Emsenhuber}
\email{emsenhuber@lpl.arizona.edu}

\author[0000-0002-8811-1914]{Alexandre Emsenhuber}
\affil{Lunar and Planetary Laboratory, University of Arizona, 1629 E. University Blvd., Tucson, AZ 85721, USA}

\author[0000-0001-6294-4523]{Saverio Cambioni}
\affil{Lunar and Planetary Laboratory, University of Arizona, 1629 E. University Blvd., Tucson, AZ 85721, USA}

\author[0000-0003-1002-2038]{Erik Asphaug}
\affil{Lunar and Planetary Laboratory, University of Arizona, 1629 E. University Blvd., Tucson, AZ 85721, USA}

\author[0000-0002-9767-4153]{Travis S. J. Gabriel}
\affil{School of Earth and Space Exploration, Arizona State University, 781 E. Terrace Mall, Tempe, AZ 85287, USA}

\author[0000-0001-5475-9379]{Stephen R. Schwartz}
\affil{Lunar and Planetary Laboratory, University of Arizona, 1629 E. University Blvd., Tucson, AZ 85721, USA}

\author{Roberto Furfaro}
\affil{Systems and Industrial Engineering Department, University of Arizona, 1127 E. James E. Rogers Way, Tucson, AZ 85721, USA}

\begin{abstract}
Terrestrial planet formation theory is at a bottleneck, with the growing realization that pairwise collisions are treated far too simply. Here, and in our companion paper \citep{2019ApJCambioni} that introduces the training methodology, we demonstrate the first application of machine learning to more realistically model the late stage of planet formation by giant impacts. We present surrogate models that give fast, reliable answers for the masses and velocities of the two largest remnants of a giant impact, as a function of the colliding masses and their impact velocity and angle, with the caveat that our training data do not yet include pre-impact rotation or variable thermal conditions. We compare canonical \textit{N}-body scenarios of terrestrial planet formation assuming perfect merger \citep{2001IcarusChambers} with our more realistic treatment that includes inefficient accretions and hit-and-run collisions. The result is a protracted tail of final events lasting $\sim \SI{200}{\myr}$, and the conversion of about half the mass of the initial population to debris. We obtain profoundly different solar system architectures, featuring a much wider range of terrestrial planet masses and enhanced compositional diversity.
\end{abstract}


% =======================
\section{Introduction}
\label{sec:intro}
% =======================

The final stage of terrestrial planetary formation is dominated by collisions between similar-sized bodies \citep{1985ScienceWetherill,2002ApJKokubo} known as giant impacts. In \textit{N}-body simulations of planet formation the usual assumption is that when planetary orbits intersect and a collision occurs, the result is a perfect merger. However, from prior studies \citep{2004ApJAgnor,2010ApJKokubo,2012ApJStewart,2018CeMDABurger} we know that a large fraction of collisions result in failed mergers known as hit-and-run collisions (HRCs). 

It has been customary to assume that if the first collision between two planets is a failed merger, they will collide again and soon achieve the merger. This has been used to justify the assumption of perfect merger in \textit{N}-body and other approaches to modeling late-stage planet formation. But this intuition is wrong. Using \textit{N}-body simulations to explicitly track both of the major remnants of an HRC (the projectile ``runner'' and the target), we have found that the bodies do \textit{not} usually collide again with each other \citep[hereafter \citetalias{2019ApJEmsenhuberA}]{2019ApJEmsenhuberA}. Indeed the runner often ends up colliding with a different planet, and so must be treated explicitly to understand its fate.

Hit-and-run is common; so is inefficient accretion, where only a fraction of the combined matter ends up in the growing planet. We need to understand the fate of the debris, and here, too, we cannot study this aspect of the problem using a code that treats collisions as perfect mergers. As we shall see, if the production of debris is taken into account, then following a long chain of planetary mergers and attempted mergers, as much as half or more of the starting mass can be potentially lost from the evolving planetary system. 

As a result, planet formation is not just a forward growth process; there are setbacks where assembled bodies can lose mass and break into smaller remnants, or where almost-merged pairs of bodies can orbit independently for another million years, perhaps never to collide again. We are motivated to apply greater realism to the planet formation process in order to better understand the specific events or chains of events that are possible and to come up with more realistic models for the growth of terrestrial planets and the architecture of the inner solar system. 

\subsection{Accretion efficiency of giant impacts}

The variable geometry of giant impacts leads to a wide range of possible outcomes as reviewed by \cite{2010ChEGAsphaug}. The relative net mass of material accreted in a given giant impact is
\begin{equation}
    \xi_\mathrm{L}=\frac{M_\mathrm{L}-M_\mathrm{T}}{M_\mathrm{P}} < 1.
    \label{eq:acclr}
\end{equation}
Here $M_\mathrm{T}$ is the mass of the target, $M_\mathrm{P}$ is the mass of the projectile, and $M_\mathrm{L}$ is the mass of the largest single gravitationally bound remnant, e.g.,\ the growing target in the case $\xi>0$. This \textit{accretion efficiency} is heavily dependent on the impact velocity relative to the mutual escape velocity $v_\mathrm{esc}$, especially in the critical range $\sim1.1-1.4 v_\mathrm{esc}$. 

Accretion efficiency also has a strong nonlinear dependence on impact angle. For impact angles of only \ang{30}, the cores of two differentiated planets might not even contact one another \citep{2010ChEGAsphaug}, leading to a radically different dynamic than a collision that is closer to head-on. 

Only the lowest-velocity giant impacts result in efficient merger, with most of those being ``graze-and-merge'' collisions (GMCs) like the canonical \ang{45} model for Moon formation at $v_\mathrm{esc}$. Even in the slowest scenarios the Earth and Theia lose several percent of their material to heliocentric orbits, for further evolution \citep{2012MNRASJacksonWyatt}. At the other extreme, collisions a few times $v_\mathrm{esc}$ are unable to accrete any mass, so $\xi<0$, blasting material from the crust and mantle of the target and obliterating the projectile. This is in the range of collisions considered for the origin of Mercury \citep{2007SSRvBenz} and the Mars hemispheric dichotomy \citep{2008NatureMarinova}.

In between---and of the greatest relevance to planetary accretion---is the highly nonlinear regime where the effect of impact angle and velocity are pronounced; here, collisions cannot be treated simply. Several numerical modeling studies of collisions have concluded that the preponderance of giant impacts result in two (or more) major remnants after the initial contact, that are either bound (GMCs; \citealp{2010ApJLeinhardt}) or unbound (HRCs; \citealp{2006NatureAsphaug}). 

As a further complicating factor, the transition between GMC and HRC (bound or unbound runner) depends not only on the impact velocity and impact angle (hit-and-run requiring faster or more grazing collisions), but also on the Hill radius, that is, the gravitational sphere of influence in the case of a massive central body \citep{2017AJRufuCanup,2019ApJEmsenhuberB}. A giant impact that happens in orbit around the sun (in the case of two planets colliding) or around a planet (in the case of two moons) tends to be more disruptive and dispersive than the same giant impact in a two-body scenario, so that certain GMCs become HRCs. 

The last stage of planet formation included the pairwise accretion of the oligarchs \citep{2002ApJKokubo}; orbital excitation was limited to the escape velocity of the most massive participants, and damped by smaller planets, so a range of collision velocities around 1.1--1.4 $v_\mathrm{esc}$ is expected \citep{2012ApJStewart,2016ApJQuintana}, with outliers at higher and lower velocities depending on damping and the presence of other major bodies. This turns out to be a quite interesting regime full of imperfect accretions and HRCs. 

Treating all these giant impacts as simple mergers certainly simplifies the modeling of accretion, but it bypasses several key phenomena of terrestrial planet formation that are relevant to their size, composition, and geologic diversity. Two aspects of recent interest are the survivor of the runner as a final planet \citep{2014NatGeoAsphaug}, and the occurrence of dynamical chains of giant impacts \citepalias{2019ApJEmsenhuberA}---one giant impact leading to another---as the rule rather than the exception. 

The late stage in any terrestrial planet-forming system therefore represents a juncture where several primary processes are happening at once: accretion of targets; disruption of projectiles; and differentiation and mixing of the contributors, frequently in collision chains with new thermal conditions that must explicitly be modeled in the presence of the other planets. 

\subsection{Effects of giant impacts}

Giant impacts have strongly varying outcomes depending on impact angle, velocity, relative mass ratio, and density stratification \citep[e.g.,][]{2010ChEGAsphaug,2019ApJGabriel}. Collisions at near-escape velocity typically result in mergers, however the impact angle $\theta_\mathrm{coll}$ is important in this scenario since this controls the amount of angular momentum in the system. Head-on collisions with near-zero angular momentum are straightforward to understand in terms of outcome, but these are rare occurrences. For most values of $\theta_\mathrm{coll}$ the result is an oblique collision where a ``runner'' emerges from the first collision with a reduced relative velocity \citepalias{2019ApJEmsenhuberA}. Hence this runner is most likely captured, and will come back for a second collision; the ensemble interaction is a GMC. It is the second event that produces the actual merger, unless deformation of the bodies leads to strong torques that can release additional material, as has been postulated for the origin of the Pluto-Charon binary \citep{2005ScienceCanup,2011AJCanup}. Oblique mergers involve the accretion of significant amount of angular momentum, and can easily form satellite systems. The ``canonical'' model of Moon formation \citep{2001NatureCanup,2004IcarusCanup} is an example. The angular momentum provided by the merger is able to change the rotation state of the target dramatically.\footnote{We would like to note that \citet{2012ScienceCuk} have also presented a mechanism for the formation of the moon. Their model, a non merging collision, is a specific case that requires the alignment of the target's spin with the impactor's trajectory, and a pre impact target that is rapidly rotating (2.4 to \SI{2.7}{\hour}) to achieve a sufficiently massive protolunar disk. In this study, however, we are interested in systematic averages, and are not accounting for the spin of the bodies.} This effect has been invoked to explain the tilted spin axis of Uranus \citep{1992IcarusSlattery,2018ApJKegerreis,2019AJKurosakiInutsuka}.

Even allowing for significant planetesimal damping, during the late stage when most mass was bound up in the largest bodies, most giant impacts occurred at velocities faster than around 1.1--1.2 $v_\mathrm{esc}$ due to gravitational stirring \citep{2010ApJKokubo,2016ApJQuintana}. In this velocity regime, for strengthless, self-gravitating bodies, the majority of giant impacts are HRCs \citep{2012ApJGenda}. For smaller bodies, the effect of friction and porosity must be considered in addition to gravity, as it will increase dissipation and enhance the cross section of coagulation \citep{2015AstIVJutzi}. Because the impact parameter $b$ lies inside the target radius, the collisional and tidal stresses strip away exterior materials during a HRC. In a differentiated body, this means that the bulk composition is modified; this is a possible origin of Mercury's anomalous core size \citep{1988IcarusBenz,2007SSRvBenz,2014NatGeoAsphaug,2018ApJChau}.

At sufficient impact angle and velocity, HRC occurs and the runner will more often than not go on to accrete with a different planet than the original target; runners in HRCs into proto-Earth, for example, are about as likely to accrete onto Venus as they are to come back to the Earth \citepalias{2019ApJEmsenhuberA}. Late-stage accretion might be more appropriately described as a chain of attempted mergers ending potentially in a merger, as opposed to simple pairwise events. It is then possible for a ``runner'' to survive multiple HRCs, which leads to compounding geological and geochemical consequences. Furthermore, the most hyperbolic (small $b$, high $v$) HRCs produce chains of compositionally-diverse runners extracted from a range of depths \citep{2013IcarusAsphaug}, that may or may not return \citep{2012MNRASJacksonWyatt}.

\subsection{A machine learning approach}

There have been a number of efforts to improve upon the assumption of perfect merger, either by applying a simple criterion in an \textit{N}-body framework \citep{2010ApJKokubo}, or using scaling laws from \citet{2012ApJLeinhardt} as in \citet{2013IcarusChambers,2016ApJQuintana,2019IcarusClement}, or by running hydrodynamical simulations on the fly during \textit{N}-body calculations \citep{2019arXivBurger}. The latter approach is most rigorous, but also the most computationally demanding; the hydrodynamic calculation requires days of compute time for appropriately high numerical resolution and has to be done for every one of the hundreds of collisions in an \textit{N}-body simulation. 

To overcome this computational bottleneck, \citet[hereafter \paperone]{2019ApJCambioni} introduced a data-driven, machine learning (ML) approach to the prediction of the outcomes of giant impacts based on a pre-existing set of simulations. Our training data is a set of hydrodynamical simulations of giant impacts for relevant sizes (\num{e-2} to \SI{1}{\mearth}) and velocities (up to 4 times the mutual escape velocity), by \citet{2011PhDReufer}. Based on the simulations we develop a `surrogate' SPH model composed by a classifier of collision types and a predictor for the mass of the largest remnant for collisions, \emph{i.e.},\ a mapping between inputs and outputs that is less constrained than a simple interpolation scheme. By training on a large data set of high-resolution simulations of Moon- to Earth-mass collisions, we are able to produce surrogate models for key outcomes of giant impacts that are accurate predictors of the outcomes needed to introduce realistic collision outcomes in an \textit{N}-body code. These parametric functions, or ``neural networks'', are able to quickly predict the outcome of a collision at a known degree of accuracy. While scaling laws \citep[e.g.,][]{2012ApJLeinhardt,2019ApJGabriel} composed of a set of algebraic functions based on physical arguments are computationally fast to query, a data-driven approach (surrogate model) does not introduce model assumptions in the fitting. Nevertheless, scaling laws provide important intuition in regards to the outcomes of giant impacts and, in the case of \citet{2019ApJGabriel}, account for density-stratification and compositional differences through physically-derived parameters, which are not covered in this study.

\subsection{This work}

Here we extend our ML approach to retrieve the mass of the two largest remnants and their relative orbit from the dataset of \paperone{}. We then develop an interface layer for the ML surrogate models (code library) that can be easily adapted to \textit{N}-body codes. The use of the new ML surrogate models enables to predict realistic post-impact trajectories of the remnants; taking into account the strong deflection of the runner's trajectory induced by the gravity of the target and the decrease of the relative velocity are taken into account. The remnants are explicitly tracked in the \textit{N}-body evolution until terminal accretion, might it be with the same target planet or with another planet \citepalias{2019ApJEmsenhuberA}.

The structure of this work is as follows: in Section~\ref{sec:dataset}, we give a brief summary of the underlying collision simulation dataset, while Sections~\ref{sec:massrad} and~\ref{sec:parameters} we provide analysis and our choice of parameters used for ML fitting. Section~\ref{sec:ml-methods} describes the relevant ML techniques and the procedure to link the ML surrogate model to \textit{N}-body algorithms. Finally, in Section~\ref{sec:ml-results}, we test our approach by reproducing and improving upon previous results of \textit{N}-body models for the formation of the solar system's terrestrial planets \citep{2001IcarusChambers,2013IcarusChambers}.

% =======================
\section{Dataset}
\label{sec:dataset}
% =======================

We use the same dataset as in \paperone, that is composed of roughly 800 simulations that were obtained by \citet{2011PhDReufer} and postprocessed in \citet{2019ApJGabriel}. The simulations were performed using the Smoothed-Particle Hydrodynamics (SPH) technique \citep[see, e.g.,][for reviews]{1992ARA&AMonaghan,2009NARRosswog}, and the (M-)ANEOS equation of state \citep{ANEOS,2007M&PSMelosh}. In the case of the solar system, we acknowledge that when objects of the mass of Earth have formed, most planetary embryos have been accreted or left the system.  However, for machine training purposes, and to extend the validity of the approach to extrasolar terrestrial planet formation, we consider the target mass $M_\mathrm{T}$ to range from \num{e-2} to \SI{1}{\mearth}. The dataset comprises 8 groups of simulations that have each one value of target mass and one value of projectile-to-target mass ratio $\gamma=M_\mathrm{P}/M_\mathrm{T}$ between 0.2 and 0.7. For each group, there are about 100 simulations with impact velocity, given in terms of the mutual escape velocity, $v_\mathrm{coll}/v_\mathrm{esc}$ between 1 and 4 and impact angles $\theta$ between 0 and \ang{90}. The mutual escape velocity is defined as:
\begin{equation}
\label{V_esc}
v_\mathrm{esc}=\sqrt{\frac{2\mu}{R_\mathrm{coll}}},
\label{eq:vesc}
\end{equation}
with $\mu=G(M_\mathrm{T}+M_\mathrm{P})$ being the standard gravitational parameter, $G$ the gravitational constant, $R_\mathrm{T}$ and $R_\mathrm{P}$ the body's radii and $R_\mathrm{coll}=R_\mathrm{T}+R_\mathrm{P}$ the separation at initial contact. The resolution of the simulation is about \num{2e5} SPH particles. While this resolution is lower than some current work, it is nevertheless sufficient to preserve the general physical processes occurring during collisions.

The training dataset is limited to 800 high resolution SPH simulations, involving $\num{2e-3}$ to \SI{1}{\mearth} differentiated ``chondritic'' (70 wt\% silicate, 30 wt\% metallic iron) planets colliding at between 1 and 4 times the mutual escape velocity, using  the ANEOS equation of state for Fe \citep{ANEOS} and M-ANEOS for SiO\textsubscript{2} \citep{2007M&PSMelosh}. The data set is best sampled for impact velocities less than $1.5 v_\mathrm{esc}$ and impact angles around \ang{45}, and is coarse at higher velocities and angles close to \ang{0} or \ang{90}. The surrogate model thus has a formal associated prediction error which is generally smaller for the well-sampled parameters. We direct the reader to \paperone{} and \citet{2019ApJGabriel} for more information about the dataset. The latter reports the entirety of the dataset that underlies our machine learning model used herein, as well as additional impact simulation datasets with various materials, e.g. water-rich bodies. It presents semi-empirical models that allow one to determine the outcomes of giant impacts quickly, in an \textit{N}-body code or otherwise, analogous to scaling laws developed for the cratering community. Furthermore, \citet{2019ApJGabriel} discusses other potential systematics in giant impacts, such as the role likely played by structure of the planet in terms of density stratification in the prediction of HRCs. That work also provides an impact velocity probability distribution based on the impacts in \textit{N}-body simulations from \citet{2013IcarusChambers} and finds that ${\sim}50\%$ of impacts occur at $\lesssim 1.2 v_\mathrm{esc}$; 95\% of the collisions occur at $<2.6 v_\mathrm{esc}$. The sampling of the dataset is adequate with respect to the expect distribution of velocities and angles.

The analysis of the final state of each simulation has been slightly updated since \paperone. The most important change is the inclusion of a friend-of-friend neighbor search to ensure that bodies escaping at near the escape velocity are correctly resolved \citepalias{2019ApJEmsenhuberA}. The analysis provides the masses of the resulting bodies, $M_\mathrm{L}$ for the largest remnant and $M_\mathrm{S}$ for the second remnant, as well as the position $\mathbf{x}$ and velocity $\mathbf{v}$ of each. In our search for $M_\mathrm{S}$, we only take the second remnant into account when $M_\mathrm{S}/M_\mathrm{P}\geq 0.1$. This criterion ensures that the second largest remnant is a HRC runner rather than debris from an erosive collision.  On the other hand, it limits what we can say, for now, about smaller remnants that would be swept up for longer time. All the values are computed at 50 times the collision time scale, defined as
\begin{equation}
\label{tg}
    \tau_\mathrm{coll} = \frac{2R_\mathrm{coll}}{v_\mathrm{coll}},
\end{equation}
where $v_\mathrm{coll}$ is the impact velocity. After this time, pressure and temperature gradient forces are no longer acting and the resulting scenario (largest remnants and their orbital properties) can be accurately treated using an \textit{N}-body integrator.

% =======================
\section{The mass-radius relationship}
\label{sec:massrad}
% =======================

\begin{figure}
	\centering
	\includegraphics{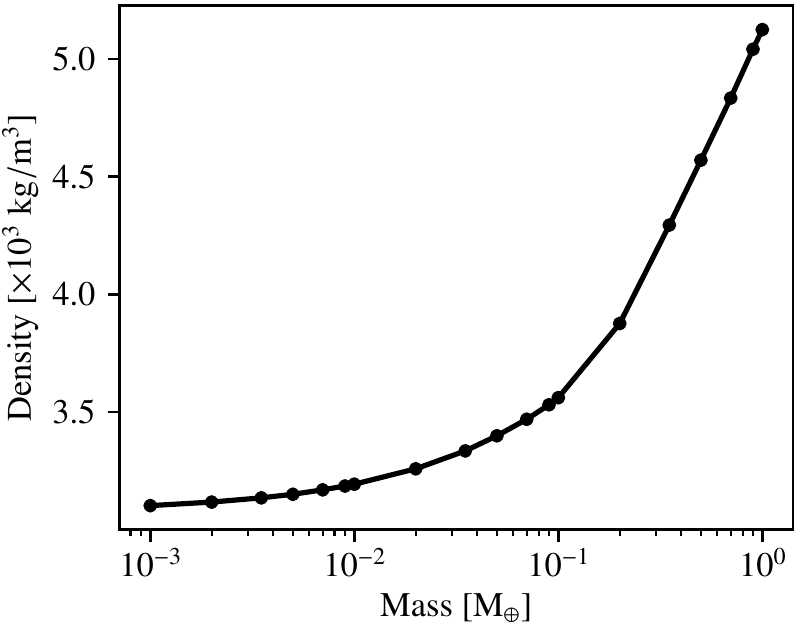}
	\caption{Bulk density as a function of the body mass, as obtained from 1D structure calculation of hydrodynamic planets using the same equation of state models as in the 3D hydrocode.}
	\label{fig:density}
\end{figure}

In addition to masses and orbits, \textit{N}-body evolution and the implementation of collisions require knowledge of the radii of the two colliding bodies. In most studies involving $N$-body evolution \citep[][e.g.]{2001IcarusChambers,2013IcarusChambers}, a constant density (and therefore mass-radius relationship) is assumed for all bodies, e.g. $\rho=\SI{3}{\gram\per\cubic\centi\meter}$, when in fact planetary materials are compressible. Reliable planetary radii are required in order to: 1) discern if a close encounter ends up being a collision, 2) compute the impact velocity relative to the mutual escape velocity (Equation~(\ref{V_esc})), and 3) compute reliable orbital parameters of the remnants.  

Our aim is to obtain a relationship consistent with the SPH simulations, and that can be applied to collisions involving planetary embryos in an \textit{N}-body scheme. This has the advantage that the radii of the resulting bodies can be directly used in the \textit{N}-body code for further evolution of the remnants, that are in turn consistent with our model should another collision occur. As the only parameter that geophysically defines the bodies in the dataset is the mass, the radius may only depend on that quantity. Hence, we assume that the resulting bodies have the same composition and thermal state as the incoming bodies. As discussed previously, neither of these is true strictly speaking, as erosion (in the case of non-grazing collisions) or material transfer during HRC predominantly affects the mantle, and supersonic collisions induce shock dissipation leading to both composition and thermal state changes. For instance, there are models to determine changes in the core mass fraction \citep{2009ApJMarcus,2010ApJMarcus}.\footnote{We chose to not address how changing the mantle fraction will affect the masses (accretion efficiency) and trajectories following a giant impact, even though the bodies in the N-body code are compositionally evolving. Our main focus is on showing differences between the common assumption of perfect merger, and what happens when inefficient accretion (e.g., hit-and-run) is realistically included.}

Further, to avoid effects that slightly change the radii depending on the numerical resolution, we derive the radius from the 1D radial profiles that we computed as part of the preparation phase following \citet{1991LNPBenz}. We show the results of this calculation for the bulk density in Figure~\ref{fig:density}. It can be noted that the densities are underestimated compared to the terrestrial planets. For instance, in our model, the Earth has a radius of $\approx\SI{6530}{\kilo\meter}$ while the literature value is about $\SI{6370}{\kilo\meter}$.

Lastly, we also need to define a relationship for masses slightly outside of the range of values provided here. For instance, the smallest mass that we can resolve is 10\% of the smallest projectile, i.e.,\ \SI{2e-4}{\mearth} while the theoretical maximum mass one object can have is a merger of the biggest target and projectile present in the dataset, i.e.,\ \SI{1.7}{\mearth} ($M_\mathrm{T}=\SI{1}{\mearth}$ with $\gamma=0.7$). To derive the radius of these bodies, we apply the following criterion: we assume that all bodies below \SI{1e-3}{\mearth} have the same bulk density, while for bodies larger than \SI{1}{\mearth}, we extrapolate the relationship given by the last two data points (\num{0.9} and \SI{1}{\mearth}), which are observed to be log-linear. For bodies less than $\SI{e-3}{\mearth}$, constant density is an appropriate approximation as internal pressure is quite limited, thus central compression is minimal. See \citet{2019ApJGabriel} for a description of the central compression of the bodies in this study. It should nevertheless be noted that this value is only valid for non-icy bodies. For instance, Ceres has a bulk density of \SI{2161.6\pm2.5}{\kilo\gram\per\cubic\meter} \citep{2019IcarusPark} but it is composed of volatiles, so it cannot be reproduced with our model.

% =======================
\section{Analysis of the dataset}
\label{sec:parameters}
% =======================

To provide all the information required to continue an \textit{N}-body integration, we need to correctly place the resulting bodies in inertial space and determine their post-collision motion. To do so, we introduce simplifying assumptions:
\begin{itemize}
	\item All motion occurs in the collision plane. We neglect deviation from this symmetry plane. Any perpendicular motion is more likely due to small inconsistencies in the preparation of the initial bodies rather to real physical effects.
	\item In case a single body remains (simple merger) we place it at the center of mass of the incoming bodies. 
	\item In case of a grazing collision resulting in two unbounded remnants (an HRC), we compute their post-impact configuration assuming a two-body problem for their relative orbit. This is then used to set the post-impact trajectories of the remnants. Effectively, the two remnants of a hit and run are directly tracked continuously in the N-body, and are therefore subject to the perturbations of all the other embryos.
\end{itemize}
Our solution requires one parameter in case a single body remains (i.e., the mass of the largest remnant), and five parameters in the case of an HRC: the mass of the bodies and three orbital elements.

\subsection{Masses}

Across the space of giant impact outcomes (accretion, erosion, and HRC regimes) a neural network is trained to predict the mass of the largest remnant $M_\mathrm{L}$ and second largest remnant $M_\mathrm{S}$. As in \paperone, we compute the resulting masses in term of a non-dimensional number, the accretion efficiency \citep{2010ChEGAsphaug}. For the largest remnant, this is Equation~(\ref{eq:acclr}) while for the second largest remnant we define
\begin{equation}
    \xi_\mathrm{S}=\frac{M_\mathrm{S}}{M_\mathrm{P}} - 1.
    \label{eq:accsr}
\end{equation}
This formulation means that $\xi_\mathrm{S}$ is almost always negative, as mass transfer from the projectile onto the target occurs during a grazing collision (\citealp{2006NatureAsphaug}; \citetalias{2019ApJEmsenhuberA}). This formulation also has the advantage that the masses of the tertiary remnants, or debris, $M_\mathrm{D}$ are well defined. If we define $\xi_\mathrm{D}=M_\mathrm{D}/M_\mathrm{P}$, then mass conservation requires that $\xi_\mathrm{L}+\xi_\mathrm{S}+\xi_\mathrm{D}=0$.

\subsection{Orbital parameters}

In the case of a HRC, we need in addition the orbital parameters of the remnants. To characterise their orbit, we chose the following: 1) the specific orbital energy $\epsilon'$, given in terms of the kinetic energy of a collision at the mutual escape velocity, 2) the impact parameter of the orbit $b'$, and 3) the shift of the longitude of the pericenter, $\Delta\varpi$. These three parameters relate to three osculating Keplerian elements: the semi-major axis $a$, the eccentric $e$ and the argument of the pericenter $\varpi$. The inclination $i$ and the longitude of the ascending node $\Omega$ are not needed due to our assumption of motion happening in a well-defined plane, while the initial anomaly will be computed by the collision resolution algorithm from the relative separation (see section~\ref{sec:collresolve-grazing}). The choice of the orbital energy rather than relative velocity is to easily allow for a later extension of the same model to GMC collisions in which the second largest remnant has a bounded orbit. In our case this parameter is defined as
\begin{equation}
\epsilon'=\frac{\Delta v'^2/2-\mu'/\Delta x'}{v'^2_\mathrm{esc}/2}=\frac{v'^2_\mathrm{coll}}{v'^2_\mathrm{esc}}-1=\frac{v'^2_\infty}{v'^2_\mathrm{esc}}
\label{eq:eorb}
\end{equation}
with $\Delta v'$ and $\Delta x'$ the relative velocity and position of the objects at the moment of analysis. The other primed quantities ($'$) are the same as the non-primed ones, expect that they are computed for the remnants $M_\mathrm{L}$ and $M_\mathrm{S}$ and their corresponding radii. The last equality in the equation is only valid in the case of an unbound orbit. The corresponding semi-major axis $a'$ can easily be retrieved from this value with
\begin{equation}
\epsilon'=-\frac{1}{2}\frac{R'_\mathrm{coll}}{a'}.
\label{eq:ascaled}
\end{equation}

Also, the choice of using the impact parameter of the resulting orbit rather than for instance the impact angle is to allow for values greater than unity. The value is calculated with
\begin{equation}
b' = \frac{h'}{R'_\mathrm{coll}v'_\mathrm{coll}}
\label{eq:para}
\end{equation}
with $h'$ the magnitude of the specific relative angular momentum $\mathbf{h}'=\mathbf{\Delta x}'\times\mathbf{\Delta v}'$.

Finally, the shift of the argument of the pericenter is calculated by the angle between the eccentricity vectors in the collision plane with
\begin{equation}
\Delta\varpi = \varpi'-\varpi.
\label{eq:shift}
\end{equation} 
We use the usual convention that $\varpi$ is measured in the same direction as the relative motion. Hence, a positive value $\Delta\varpi$ indicates that argument of the pericenter is shifted forward during the collision, as viewed by the projectile.

 \begin{table*}
    \centering
    \caption{Excerpt of the data from the collision simulation analysis.}
    \begin{tabular}{cccc|c|ccccc}
        \hline
        Target mass & Mass ratio & Angle & Velocity & Type & Acc. L. R. & Acc. S. R. & Energy & Parameter & Shift of peri. \\
        $M_\mathrm{T}\ [\si{\mearth}]$ & $\gamma=M_\mathrm{P}/M_\mathrm{T}$ & $\theta$ [deg] & $v_\mathrm{coll}/v_\mathrm{esc}$ & & $\xi_\mathrm{L}$ & $\xi_\mathrm{S}$ & $\epsilon'$ & $b'$ & $\Delta\varpi$ [rad] \\
        \hline
        \hline
        \num{1}   & 0.70 & 52.5 & 1.15 &  1 &  0.02 & -0.03 & 0.04 & 0.76 & -0.63 \\
        \num{1}   & 0.70 & 22.5 & 3.00 &  1 & -0.58 & -0.62 & 3.72 & 0.44 & -0.62 \\
        \num{1}   & 0.70 & 45.0 & 1.30 &  1 &  0.02 & -0.04 & 0.26 & 0.80 & -0.62 \\
        \num{e-1} & 0.70 & 15.0 & 1.40 &  0 &  0.90 & -1.00 &  ... &  ... &   ... \\
        \num{e-1} & 0.20 & 15.0 & 3.50 & -1 & -1.51 & -1.00 &  ... &  ... &   ... \\
        \num{e-1} & 0.35 & 15.0 & 3.50 & -1 & -1.25 & -1.00 &  ... &  ... &   ... \\
        \num{e-2} & 0.70 & 60.0 & 1.70 &  1 &  0.00 & -0.02 & 1.61 & 0.88 & -0.14 \\
        \hline
    \end{tabular}
    \tablecomments{This table is published in its entirety in the machine-readable format. A portion is shown here for guidance regarding its form and content.}
    \tablecomments{The elements in the first four columns are the predictors (collision properties): $M_T \in [\num{e-2},~1]~\si{\mearth}$; $\gamma=M_\mathrm{P}/M_\mathrm{T} \in [0.2,~0.7]$; $\theta_{coll} \in [0,~\ang{90}]$; $v_{coll}/v_{esc} \in [1,~4]$. The elements in the other columns are the responses (type of collision outcome). The fifth column (Type) is the automated classification (for the classifier), with HRCs coded as \#1, accretion as \#0, and erosion as \#-1. The last five columns are the responses for the NNs: $\xi_\mathrm{L}$ is computed according to Equation~(\ref{eq:acclr}), $\xi_\mathrm{S}$ to Equation~(\ref{eq:accsr}), $\epsilon'$ to Equation~(\ref{eq:eorb}), $b'$ to Equation~(\ref{eq:para}) and $\Delta\varpi$ to Equation~(\ref{eq:shift}).}
    \label{tab:data}
\end{table*}

% =======================
\section{ML techniques}
\label{sec:ml-methods}
% =======================

We use ML techniques to train a classifier of collision type and a regressor for collision outcomes. The training, validation and testing procedures are analogous to those described in \paperone, to which we refer for more details. In the following, we describe the differences in the labelling process of the data. An excerpt of the labelled data is reported in Table \ref{tab:data}. The dataset is published in its entirety in the machine-readable format.

% ----------------------------------------
\subsection{Classifier of collision types}
\label{sec:classifier}
% ----------------------------------------

The classifier presented in \paperone{} is here revised by training on type of collisions (classes, or responses) based on the following mass criterion: accretion ($M_\mathrm{L}>M_\mathrm{T}$ and $M_\mathrm{S}<0.1 M_\mathrm{P}$), erosion ($M_\mathrm{L}<M_\mathrm{T}$ and $M_\mathrm{S}<0.1 M_\mathrm{P}$), and HRCs ($M_\mathrm{S}<0.1 M_\mathrm{P}$). The outcome of the simulations is associated to four impact parameters (predictors): mass of the target, projectile-to-target mass ratio, impact angle, and impact velocity. The dataset has entries:
\begin{equation}
    \{(M_\mathrm{T},~\gamma,~\theta_\mathrm{coll},~v_\mathrm{coll}/v_\mathrm{esc}) ; \mathrm{class}\}.
\end{equation}

We quantify the performance of the classifier in terms of its confusion matrix at testing. As described in more detail in \paperone, the confusion matrix shows the degree to which the classifier is confused when it makes predictions; each row represents the instances in a predicted class while each column represents the instances in an actual class \citep{Ting2010}. The accuracy of the classifier is computed as the percentages of true positives (correct predictions) over total number of sample classification problems. A number of classes greater than 2 are decomposed into multiple binary classification problems, according to different transformation techniques \citep[e.g., one-vs-one and one-vs-rest strategies,][]{Bishop2006}. The choice of a specific technique is also part of the process of hyperparameter optimization, which consists in choosing the combination of non-learned features (e.g., the type of algorithm and size of the cross-validation batch) that maximizes the accuracy of the trained scheme.

\subsection{Regressor of collision outcomes}

We design a surrogate model which predicts the outcome of the collision according to four impact parameters (predictors): mass of the target, projectile-to-target mass ratio, impact angle, and impact velocity. With the term ``surrogate model" we mean a parametric function trained to mimic the ``parent'' SPH input-output function to predict real-variable outputs given the input parameters (predictors). Running the surrogate model drastically reduces the computational time with respect to full SPH simulations (from hours to seconds), thus enabling `on-the-fly' predictions of planetary collision outcomes during \textit{N}-body planetary formation studies.

For this surrogate model, the dataset has the following entries:
\begin{equation}
    \{(M_\mathrm{T},~\gamma,~\theta_\mathrm{coll},~v_\mathrm{coll}/v_\mathrm{esc}) ; (\xi_\mathrm{L}, \xi_\mathrm{S})\}.
\end{equation}

In the HRC regime only, a second neural network is trained to predict the post-collision hyperbolic orbit of the second largest remnant with respect to the largest remnant. Specifically, the predicted outcomes are: 1) the specific orbital energy $\epsilon'$, given in terms of the kinetic energy of a collision at the mutual escape velocity, 2) the impact parameter of the orbit $b'$, and 3) the shift of the longitude of the pericenter, $\Delta\varpi$. For this surrogate model, the dataset has entries:
\begin{equation}
    \{(M_\mathrm{T},~\gamma,~\theta_\mathrm{coll},~v_\mathrm{coll}/v_\mathrm{esc}) ; (\epsilon', b', \Delta\varpi)\}.
\end{equation}
As we demonstrate in section \ref{sec:collresolve}, the predicted quantities are sufficient to fully describe the post-collision geophysical and orbital state of the remnants and hence to further evolve the remnant(s) in the \textit{N}-body study. 

Analogously to \paperone, the training of the networks is performed on 70\% of the overall dataset. The rest of the data is split between a validation set (15\%) and a testing set (15\%). The dataset is split via random sampling without replacement to assure that the data in the three sets follow the same probability distribution. The networks are optimized by tuning the following hyperparameters: number of hidden layers, number of hidden units, batch normalization technique, activation function, regularization strength and training technique (including relevant parameters, e.g., learning rate). The combination of hyperparameters that achieves the best performance at validation is chosen to define the architecture of the network. After training, the performance of the networks are assessed by means of the Mean Square Error (MSE) and correlation index at testing. The MSE value is an estimate of the global error, as it quantifies the (squared) residual between the values predicted by the regressor and the values of the outcome of the SPH data (e.g., accretion efficiency) in the testing set. The correlation index R measures the degree of correlation between outputs and targets; this quantity is the analogous of the SVM classification accuracy for real-variable data. An optimal result shows low MSE values (i.e., close to zero) and a high degree of correlation between predictions and targets (i.e., a R value close to 100\%).

% =======================
\section{ML Results}
\label{sec:ml-results}
% =======================

% =======================
\subsection{Performance: classifier of collision types}
\label{sec:class_perf}
% =======================

\begin{figure}
	\centering
	\includegraphics[width=\linewidth]{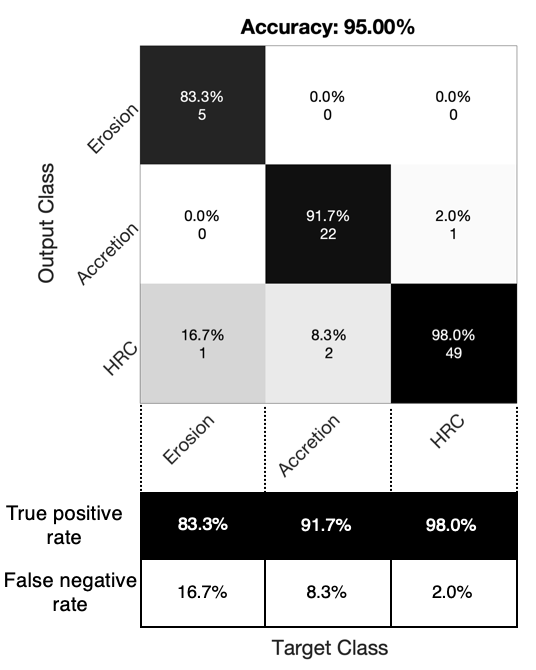}
	\caption{Confusion matrix of the 3-D classifier, quantifying the degree of accuracy of the classification on the testing set. The elements on the diagonal of the confusion matrix represents those instances that have been correctly classified by the SVM (true positives). Conversely, each extra-diagonal element represents the number of mis-classifications with respect the SPH data (i.e., the labels). The number of misclassifications is added along each column to compute the false negative rates. Overall, we achieve a true positive rate of 83.3\% on the ``erosion'' class, 91.7\% for ``accretion'' class, 98.0\% for the HRC class. The confusion matrix is close to be fully diagonal; the accuracy -- computed as the mean value of the true positives over the whole population-- is 95\%.}
	\label{fig:confusion}
\end{figure}

Among the available schemes, we selected a multi-class Support Vector Machine \citep[SVM,][]{hearst1998support} as the algorithm achieving the highest validation accuracy for the classification task (93.1\%) with a size of cross-validation batch equal to 10\%. The performance of the trained classifier is quantified by the confusion matrix in Figure~\ref{fig:confusion}. The matrix is computed on a testing set corresponding to 80 entries (10\% of the total dataset), which was not used for training and cross-validation. We achieve an overall accuracy equal to 95\% at testing.

% =======================
\subsection{Performance: surrogate collision model}
\label{sec:reg_perf}
% =======================

\begin{figure*}
	\centering
	\includegraphics[width=\linewidth]{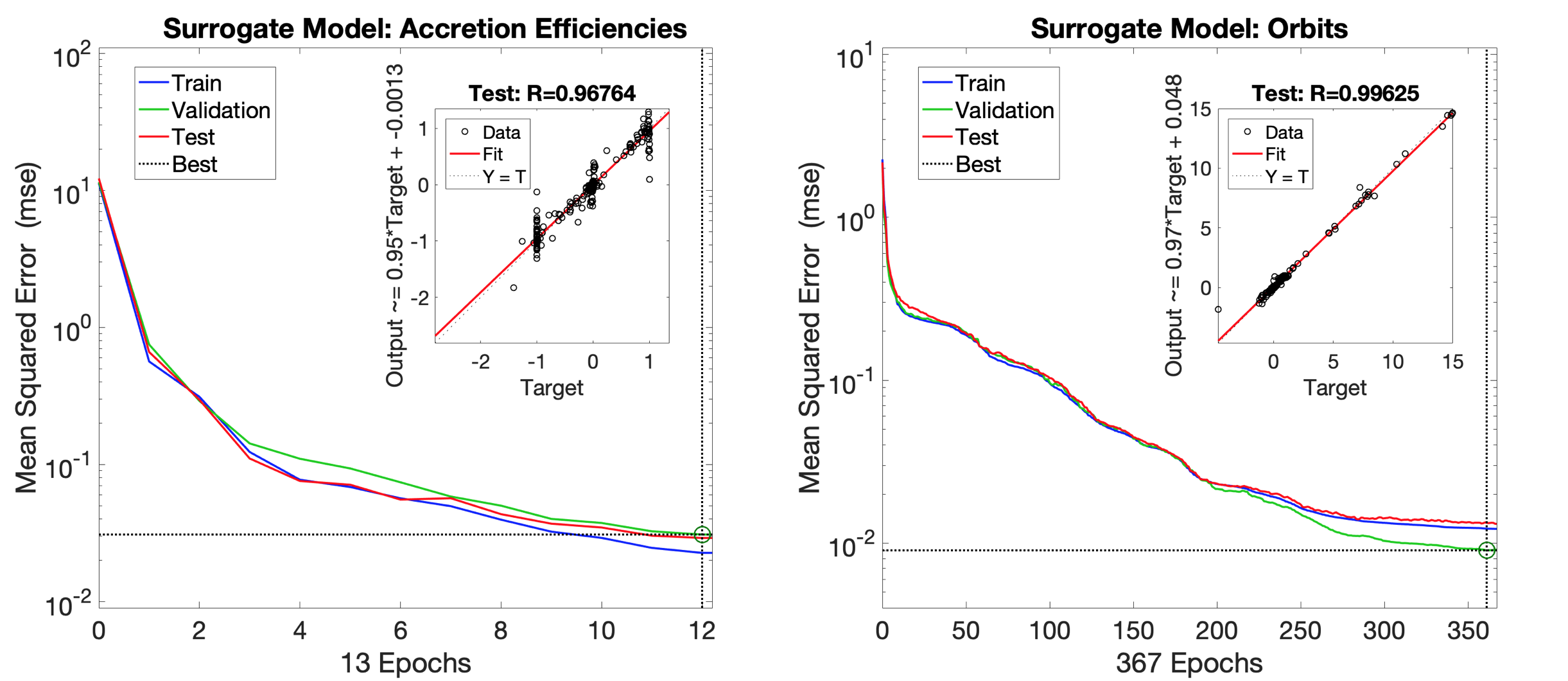}
	\caption{Performance of the  neural networks in terms of Mean Square Error (MSE), which quantifies the global uncertainty of the surrogate model in mimicking the ``parent'' numerical model, i.e., the SPH simulations, and regression index. Left panel: accretion efficiency regressor (at testing, MSE: 0.03; $R>96\%$). Right Panel: orbital regressor (at testing, MSE: 0.01; $R>99\%$).}
	\label{fig:reg_perf}
\end{figure*}

The neural network architectures that we use in the \texttt{collresolve} library are described in Table~\ref{tab:NNarch}. Such combination of hyperparameters is chosen because it minimizes the validation MSE during hyperparameter tuning. Figure~\ref{fig:reg_perf} shows learning dynamics in terms of the evolution of the MSE for training, validation and testing, at different epochs of training procedure, for both the accretion and orbital parts of the surrogate models (left and right panel, respectively). The box plots show the regression index at testing. The training, validation and testing MSE of the accretion efficiency neural network converges to a value of about 0.03. The correlation index at the end of the training procedure is above 96\% on testing. For the performance of the orbital neural network, the testing, training and validation MSE converge to an error level of about 0.01, and the correlation index is above 99\% on testing. 

\begin{table*}
    \centering
    \caption{Choice of hyperparameters for the neural networks, achieving the best performances on the validation set.}
    \begin{tabular}{|c|c|c|c|c|}
    \hline
        Network & Training Algorithm & Hidden Layers and Activation Functions & Normalization  & Regularization \\ \hline
        Accretion & Levemberg-Marquardt & Neurons: (10,~7); Activation: tansig & MinMax & $1\times10^{-3}$\\
        Orbital & Gradient Descent ($\alpha$ = 1)& Neurons: (50,~45,~15); Activation: tansig & Gaussian & $9\times10^{-1}$\\
        \hline
    \end{tabular}
    \tablecomments{The training algorithms (Levemberg - Marquardt and Gradient Descent) are described in \citet[]{demuth2014neural} and \citet[]{bottou2010large}, respectively. The term ``tansig" is an abbreviation for the hyperbolic tangent sigmoid transfer function (Eq. 13 in \paperone). The column ``normalization" refers to the common practice of scaling the inputs and targets so that they fall within a specified range \citep{ioffe2015batch}. The procedure ``minmax'' transform inputs and targets in the range [-1,1]. The procedure ``Gaussian'' normalizes the data for the mean and standard deviation of the training set. The column ``regularization'' reports on the strength of the regularization process \citep[e.g.,][for more details]{girosi1995regularization}.}
    \label{tab:NNarch}
\end{table*}

% =======================
\subsection{Consistency analysis: classifier vs regressor}
\label{sec:ml}
% =======================

\begin{figure*}
	\centering
	\includegraphics{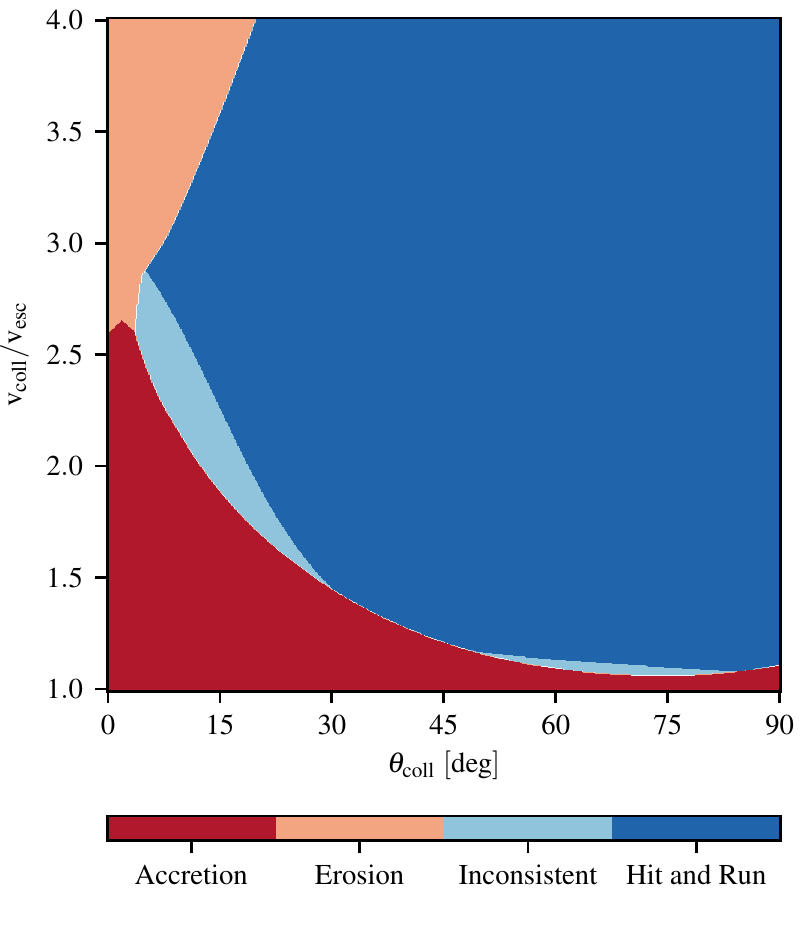}
	\includegraphics{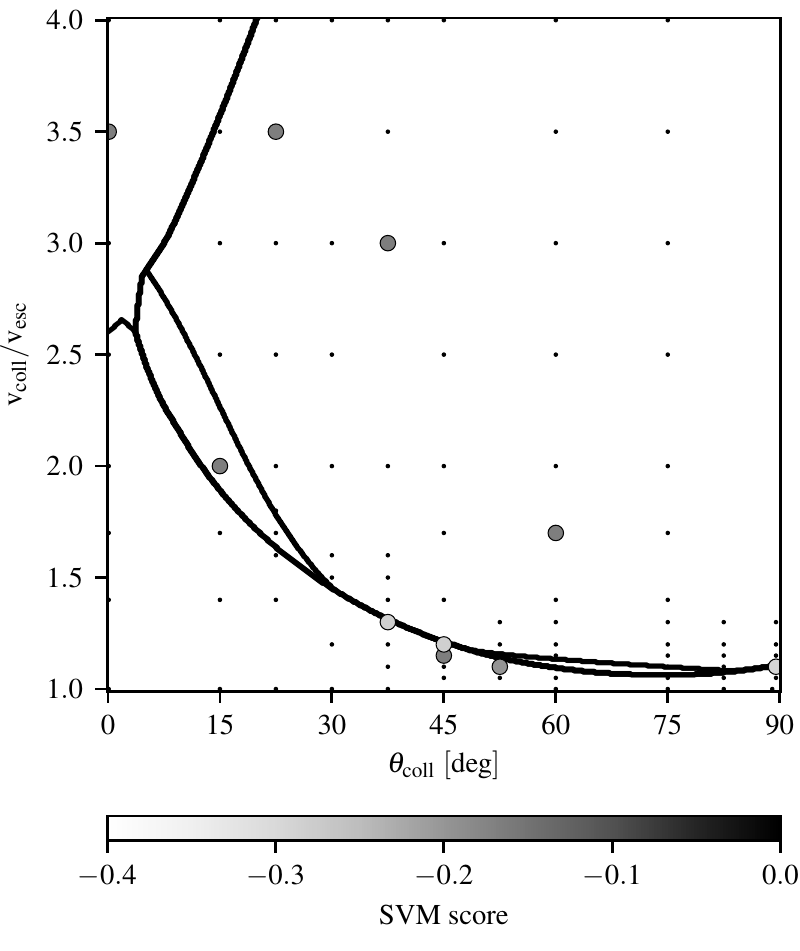}
	\caption{Decision boundaries for the classifier, coupled check for the specific orbital energy, as function of the impact angle and velocity, for a target mass $M_\mathrm{T}=\SI{0.1}{\mearth}$ and projectile mass ratio $\gamma=0.7$. Colors are as follow: red is for accretion with a single remnant ($M_\mathrm{L}>M_\mathrm{T}$ and $M_\mathrm{S}<0.1 M_\mathrm{P}$), pink for erosion ($M_\mathrm{L}<M_\mathrm{T}$ and $M_\mathrm{S}<0.1 M_\mathrm{P}$), dark blue for HRC ($M_\mathrm{S}>0.1 M_\mathrm{P}$) and light blue for when the classifier determines it is a HRC, but the orbital energy is negative.}
	\label{fig:classes}
\end{figure*}

\begin{figure*}
	\centering
	\includegraphics{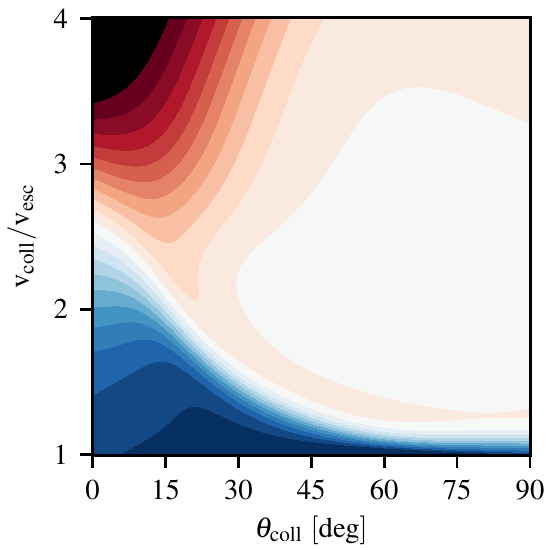}
	\includegraphics{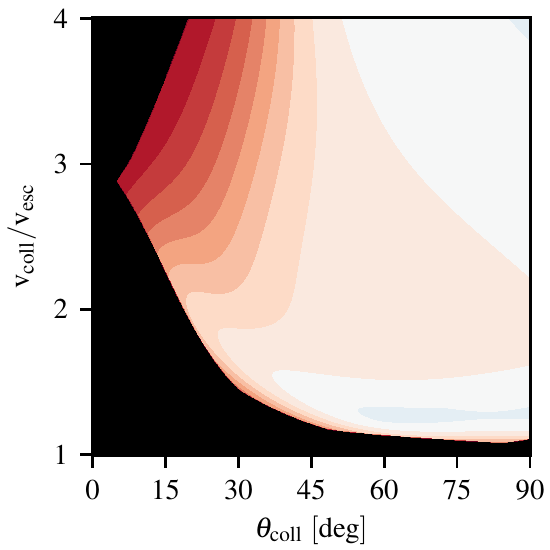}
	\includegraphics{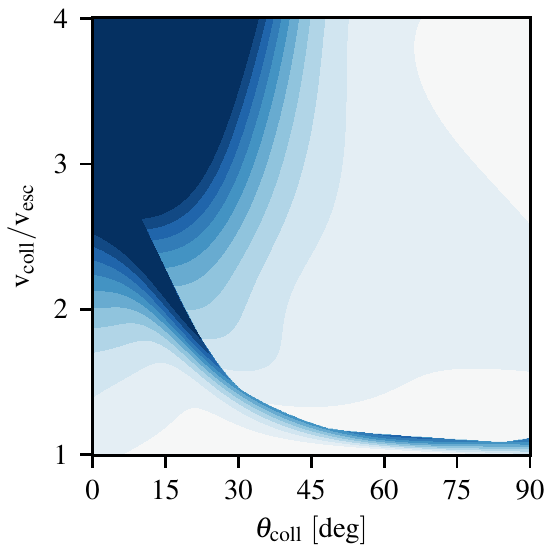}
	\includegraphics{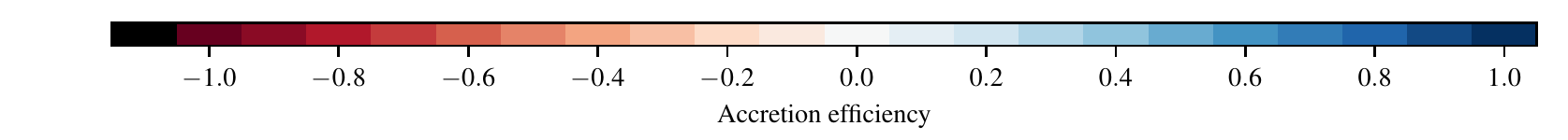}
	\caption{Left: accretion efficiency for the largest remnant $\xi_\mathrm{L}$, Equation~(\ref{eq:acclr}); Center: accretion efficiency for the second remnant $\xi_\mathrm{S}$, Equation~(\ref{eq:accsr}); Right: mass of debris $\xi_\mathrm{D}$; all as function of the impact angle and velocity, for a target mass $M_\mathrm{T}=\SI{0.1}{\mearth}$ and projectile mass ratio $\gamma=0.7$.}
	\label{fig:acc}
\end{figure*}

\begin{figure*}
	\centering
	\includegraphics{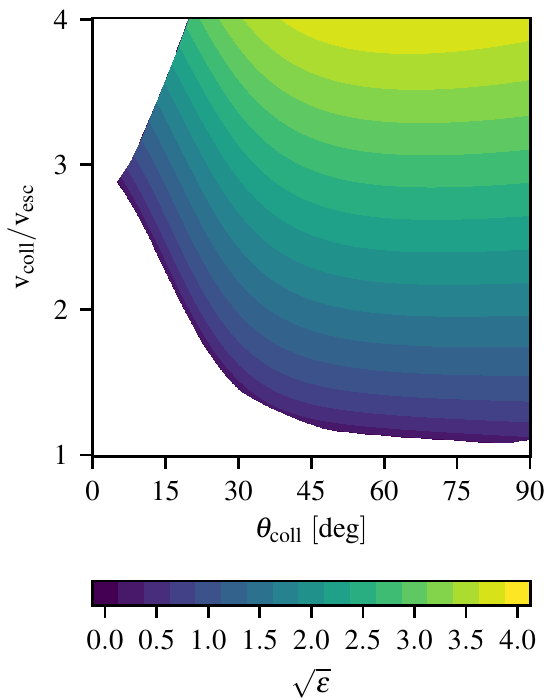}
	\includegraphics{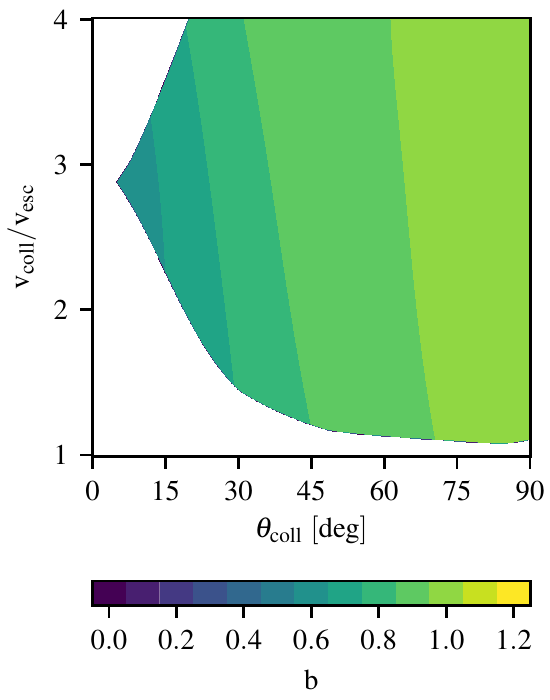}
	\includegraphics{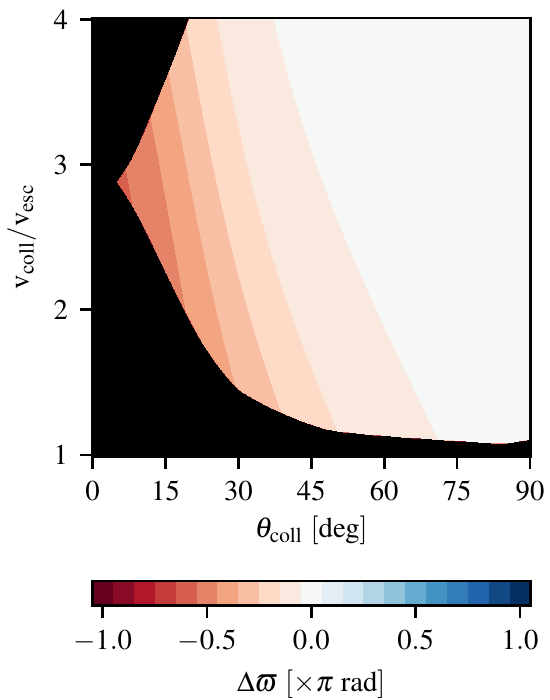}
	\caption{Orbital parameters for the case with two remnants. Left: scaled orbital energy; center: impact parameter; right: shift of the direction of the pericenter; all as function of the impact angle and velocity, for a target mass $M_\mathrm{T}=\SI{0.1}{\mearth}$ and projectile mass ratio $\gamma=0.7$. The white (for the two left panels) or black (in the right panel) region is where only a single significant remnant is produced and hence where these parameters do not have any meaning.}
	\label{fig:orb}
\end{figure*}

In Figure~\ref{fig:classes}, left panel, we show the boundary decision of the classifier for a target mass $M_\mathrm{T}=\SI{0.1}{\mearth}$ and projectile mass ratio $\gamma=0.7$. This slice of the parameter space is compared with the result from the surrogate model in terms of the relative orbit in the case of HRC. We note that there are two locations where the classifier determines that collision is in the HRC regime, while the surrogate model for the orbit provides a negative value for the orbital energy, meaning that the orbit is bound (in light blue in the figure). This regions is, hence, where the two algorithms give different answers. However, the size of this region is small, and close to the transition between the accretion (whether non-grazing or GMC) and HRC regimes. As a result, although the two answers above are different, they are close to each other. That is, despite their different methodologies, the two different algorithms produce relatively similar results.

The classifier and the regressor of orbital energy provide a consistent answer within the band of uncertainty of the former (Figure~\ref{fig:classes}, right panel). While the classifier has high global accuracy, certain regimes are characterized by more misclassifications, which prevent the classifier from achieving 100\% accuracy at testing (i.e., a fully diagonal confusion matrix). Also contributing to misclassifications are outliers in the SPH simulation dataset. For instance, three misclassifications appears in the HRC regime on the right panel of Figure~\ref{fig:classes}; these are labeled as erosive events. This suggests that some phenomenon occurred in those specific SPH simulations resulting in nonphysical outcome. On the other hand, the presence of these outliers shows a strength of the ML approach: the area surrounding these events is still classified as part of the HRC regime. Hence, the presence of a limited amount (less than 10\%) of outliers does not affect significantly the decision boundaries of the classifier.

Also we provide the mass of the remnant, given in terms of the projectile mass in Figure~\ref{fig:acc}, and for the cases where two remnants get out of the collision, the orbital parameters of their relative orbit Figure~\ref{fig:orb}.

The first orbital element is provided as $\sqrt{\epsilon'}=v'_\infty/v'_\mathrm{esc}$. It can be noted that the value close to the boundary with the accretion regime has values close to zero, which is consistent with the GMC regime. On the other hand, the transition with the erosion regime has no such constraint, as this part is due to the low mass of the second remnant. Further away from the transition, the value is mainly constrained by the relative velocity of the incoming bodies. The second parameter is the impact parameter of the second collision, $b'$.

The classifier and surrogate models that are presented here are available as a part of the \texttt{collresolve} library that is introduced in the next section.

% =======================
\section{Procedure to resolve a collision}
\label{sec:collresolve}
% =======================

\begin{figure}
	\centering
	\includegraphics{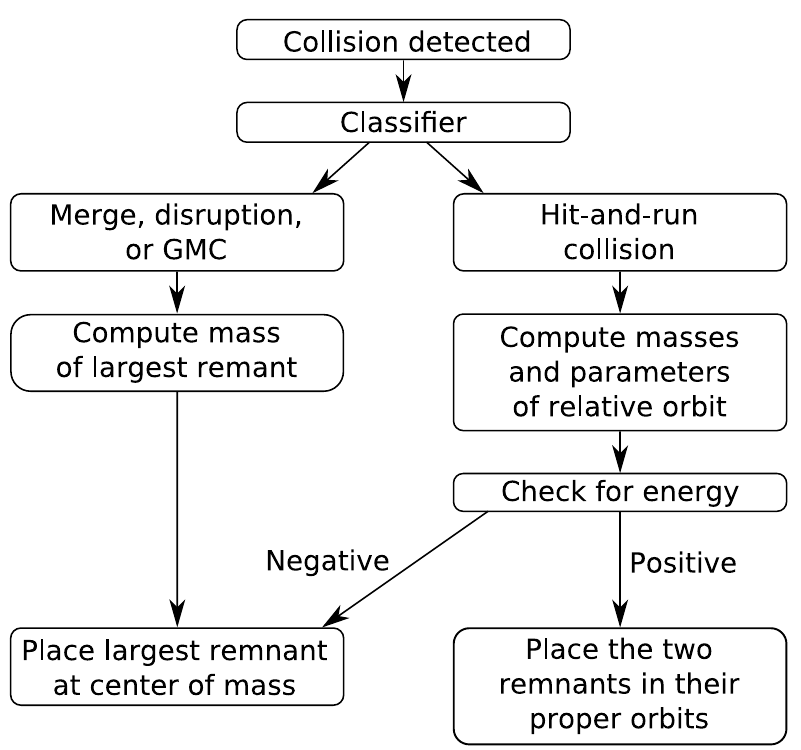}
	\caption{Sketch of the procedure being used to determine the remnant bodies when a collision has been detected.}
	\label{fig:flow}
\end{figure}

To determine the outcome and the properties of the resulting bodies when a collision occurs during an \textit{N}-body simulation, we employ the procedure that is outlined in this section. Our approach has been implemented in a code library named \texttt{collresolve}\footnote{\url{https://github.com/aemsenhuber/collresolve}} \citep{2019SoftwareEmsenhuberCambioni}. This library was designed to be easily adapted to  \textit{N}-body codes. The library can be imported into an \textit{N}-body code and used as a collision resolver. The library itself is written in \texttt{C} and has an interface for the \texttt{C}, \texttt{C++}, \texttt{Fortran} and the Python Programming Language. Examples of implementation in \textit{N}-body codes are provided along with the library.

When the \textit{N}-body code flags two objects as collided, the following inputs must be provided to the \texttt{collresolve} library: the mass $M$, radius $R$, position $x$ and velocity $v$ of the two colliding bodies, the target and projectile, with indices ``T'' and ``P'' respectively, where $M_\mathrm{T}\geq M_\mathrm{P}$. The library operates an internal conversion of the inputs into the four valued required by the ML algorithms: the mass of the target, the projectile-to-target mass ratio, the impact angle, and the velocity. These are provided to the ML algorithms according to the logical flowchart outlined in Figure~\ref{fig:flow}. The output of the library includes the mass of the remnants (largest and second largest, according to the type of collision) and their relative orbits, in a format that allows to further evolve the remnants in the \textit{N}-body code, plus the mass of unaccreted or stripped debris.

For now, we treat the debris as a single component and track only its mass. We also assume that the bodies are not spinning prior to the collision and not spinning afterwards. However we acknowledge that this violates the conservation of angular momentum in the collision. We note that follow-up collisions between the runner and target of a HRC will not have any preferential geometry with respect to the first collision which may mitigate the detrimental effects of neglecting pre-impact spin \citepalias{2019ApJEmsenhuberA}, though we acknowledge this as an important area of future work.

% --------------------------------
\subsection{Collision parameters}
% --------------------------------

As the initial step, \texttt{collresolve} computes the collision parameters that are the inputs of our surrogate models and the classifier. The target mass $M_\mathrm{T}$ and mass ratio $\gamma=M_\mathrm{P}/M_\mathrm{T}$ are readily available. To obtain the other two parameters, the impact angle $\theta$ and impact velocity $v_\mathrm{coll}$, we first compute the specific orbital energy
\begin{equation}
	\epsilon_\mathrm{orb} = \frac{\mathbf{\Delta v}^2}{2}-\frac{\mu}{\Delta x},
\end{equation}
where $\epsilon_\mathrm{orb}>0$ for the general case of colliding independent bodies,  
and specific relative angular momentum
\begin{equation}
\mathbf{h}=\mathbf{\Delta x}\times\mathbf{\Delta v},
\end{equation}
with $\mu=G(M_\mathrm{T}+M_\mathrm{P})$ the standard gravitational parameter of the incoming bodies, and $\mathbf{\Delta x}=\mathbf{x}_\mathrm{P}-\mathbf{x}_\mathrm{T}$ and $\mathbf{\Delta v}=\mathbf{v}_\mathrm{P}-\mathbf{v}_\mathrm{T}$ their relative position and velocity. The angular momentum vector determines the collision plane. Then, the impact velocity is
\begin{equation}
v^2_\mathrm{coll}=2\left(\epsilon_\mathrm{orb}+\frac{\mu}{R_\mathrm{coll}}\right)
\label{eq:vcoll}
\end{equation}
and the impact parameter
\begin{equation}
\sin{\theta}=b=\frac{h}{R_\mathrm{coll}v_\mathrm{coll}}.
\end{equation}
This formulation ensures that the values are consistent with the SPH simulation, as \textit{N}-body codes do not usually provide the values exactly at initial contact. We also include a check that $b\leq 1$, otherwise this cannot be treated as a collision. This second check is needed for some \textit{N}-body codes. In \textit{Mercury} for instance, the minimum separation is determined using a cubic polynomial fitting of the relative positions and velocities at the beginning and the end of a step. We observed that this approach can cause bodies to be flagged as having collided while our method returns $b>1$. We acknowledge that this situation may still be a mantle-stripping tidal event \citep[e.g.,][]{2006NatureAsphaug} but this is beyond the predictive capabilities of our surrogate model.

The second step is the classification of the type of collision. As described in Section \ref{sec:classifier}, the ML classifier declares the collision as accretionary, erosive or HRC. According to the type, the collision will be further handled as an accretion problem with a single remnant (in the case of merger, disruption or GMC, left branch of the chart in Figure~\ref{fig:flow}) or a grazing collision with two remnants that can be directly tracked (in the case of a HRC, right branch of the chart in Figure~\ref{fig:flow}). 

\subsubsection{Single remnant}

In the case of a collision resulting in a single major remnant, that body is given the mass $M_\mathrm{L}$ and emplaced at the center of mass of the collision, which is given by
\begin{eqnarray}
\mathbf{x}_\mathrm{COM} & = & \frac{1}{1+\gamma}\left(\mathbf{x}_\mathrm{T}+\gamma\mathbf{x}_\mathrm{P}\right), \\
\mathbf{v}_\mathrm{COM} & = & \frac{1}{1+\gamma}\left(\mathbf{v}_\mathrm{T}+\gamma\mathbf{v}_\mathrm{P}\right).
\end{eqnarray}

The remaining lost mass is provided as debris in $M_\mathrm{D}$ and its use is left to the caller. The mass of the second remnant $M_\mathrm{S}$ is not determined.

\subsubsection{Grazing collision}
\label{sec:collresolve-grazing}

For an HRC, the orbital part of the surrogate model provides the values $(\epsilon',b',\Delta\varpi)$. As a consistency check, we ensure that the obtained value of $\epsilon'$ is positive. If this is not the case, then we treat the collision in the same manner as non-grazing events. The orbital elements determine the relative motion of the remnants along the center of mass of the incoming bodies. In addition to those, the initial relative separation of the bodies is parameterized by $d=\Delta x'/R'_\mathrm{coll}$, which depends on the possibilities of the \textit{N}-body scheme and its method to determine collisions. To obtain the relative orbit of the two remnants, the first step is to retrieve the eccentricity vector
\begin{equation}
\mathbf{e}=\frac{\mathbf{\Delta v}\times\mathbf{h}}{\mu}-\frac{\mathbf{\Delta x}}{\Delta x}.
\end{equation}
With this value determined, the direction of the new eccentricity vector $\hat{\mathbf{e}}'=\mathbf{e}'/e'$ is obtained by making a rotation of $\hat{\mathbf{e}}=\mathbf{e}/e$ about $\hat{\mathbf{h}}=\mathbf{h}/h$ by an angle of $\Delta\varpi$. Thus, the orientation of the new orbit is given by $\hat{\mathbf{e}}'$ for the direction of the major axis and $\hat{\mathbf{q}}'=\hat{\mathbf{h}}\times\hat{\mathbf{e}}'$ for the minor axis. Then, to compute the other necessary quantities, we remember the relationship for the specific orbital energy,
\begin{equation}
    \epsilon'_\mathrm{orb}=-\frac{\mu'}{2a'}=\frac{\mu'}{R'_\mathrm{coll}}\epsilon'
    \label{eq:orbenergy}
\end{equation}
where we used Equation~(\ref{eq:ascaled}) for the second equality and
\begin{equation}
    p'=\frac{h'^2}{\mu'}=a'(1-e'^2)
    \label{eq:latusrectum}
\end{equation}
with $p'$ being the semi-latus rectum (see, e.g., Equations~(2.16) and~(2.17) of \citealp{2000BookMurrayDermott} or Equation~(4.18) and Table~4.1 of \citealt{2005BookBeutler1}).
Using the first equality of Equation~(\ref{eq:latusrectum}), substituting $h'$ using Equation~(\ref{eq:para}), and using $v'^2_\mathrm{coll}=2(\mu'/R'_\mathrm{coll})(1+\epsilon')$ from the combination of Equation~(\ref{eq:vcoll}) and the second equality of Equation~(\ref{eq:orbenergy}), we obtain
\begin{equation}
\frac{p'}{R'_\mathrm{coll}} = 2b'^2(1+\epsilon').
\end{equation}
With the second equality of Equation~(\ref{eq:latusrectum}), we obtain $e'^2=1+2\epsilon'_\mathrm{orb}p'/\mu'$, hence
\begin{equation}
e'^2 = 1 + 4b'^2\epsilon'(1+\epsilon').
\end{equation}
Finally, the parameter $d$ is linked to the true anomaly at initial contact with
\begin{equation}
\cos{f}=\frac{1}{e'}\left(\frac{p}{dR'_\mathrm{coll}}-1\right),
\end{equation}
and $\sin{f}=\sqrt{1-\cos^2{f}}$ as the bodies are moving away from each other. The relative position and velocity are thus
\begin{eqnarray}
\mathbf{\Delta x}' & = & \Delta x'\left(\hat{\mathbf{e}}'\cos{f}+\hat{\mathbf{q}}'\sin{f}\right) \\
\mathbf{\Delta v}' & = & \sqrt{\frac{\mu'}{p'}}\left(\hat{\mathbf{e}}'(-\sin{f})+\hat{\mathbf{q}}'(e'+\cos{f})\right),
\end{eqnarray}
which leads to the absolute position and velocity of the resulting bodies
\begin{eqnarray}
\mathbf{x}_\mathrm{L} & = & \mathbf{x}_\mathrm{COM} -\frac{\gamma'}{1+\gamma'} \mathbf{\Delta x}' \\ \mathbf{x}_\mathrm{S} & = & \mathbf{x}_\mathrm{COM} +\frac{1}{1+\gamma'} \mathbf{\Delta x}' \\
\mathbf{v}_\mathrm{L} & = & \mathbf{v}_\mathrm{COM} -\frac{\gamma'}{1+\gamma'} \mathbf{\Delta v}' \\ \mathbf{v}_\mathrm{S} & = & \mathbf{v}_\mathrm{COM} +\frac{1}{1+\gamma'} \mathbf{\Delta v}'.
\end{eqnarray}

% =======================
\section{Application to the formation of the Solar system's terrestrial planets}
\label{sec:nbody}
% =======================

\begin{figure*}
	\centering
	\includegraphics{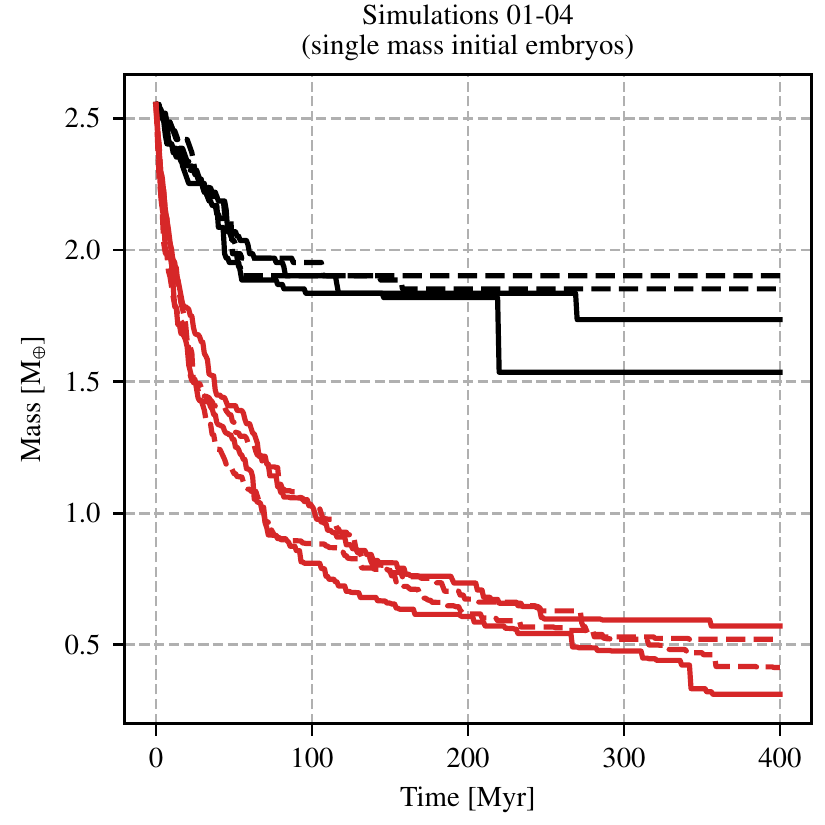}
	\includegraphics{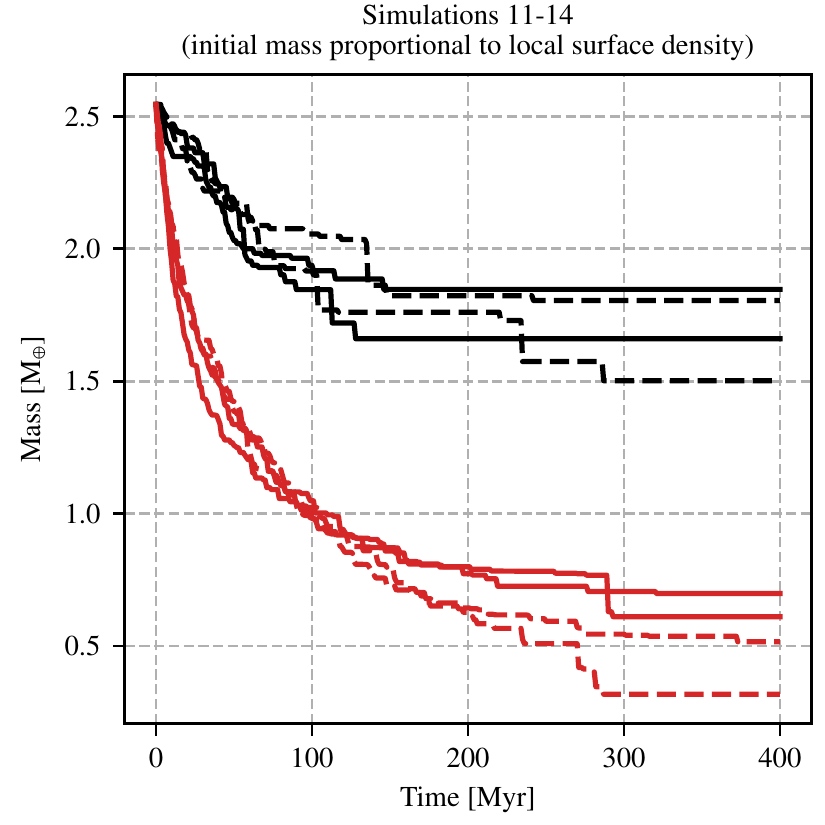}
	\includegraphics{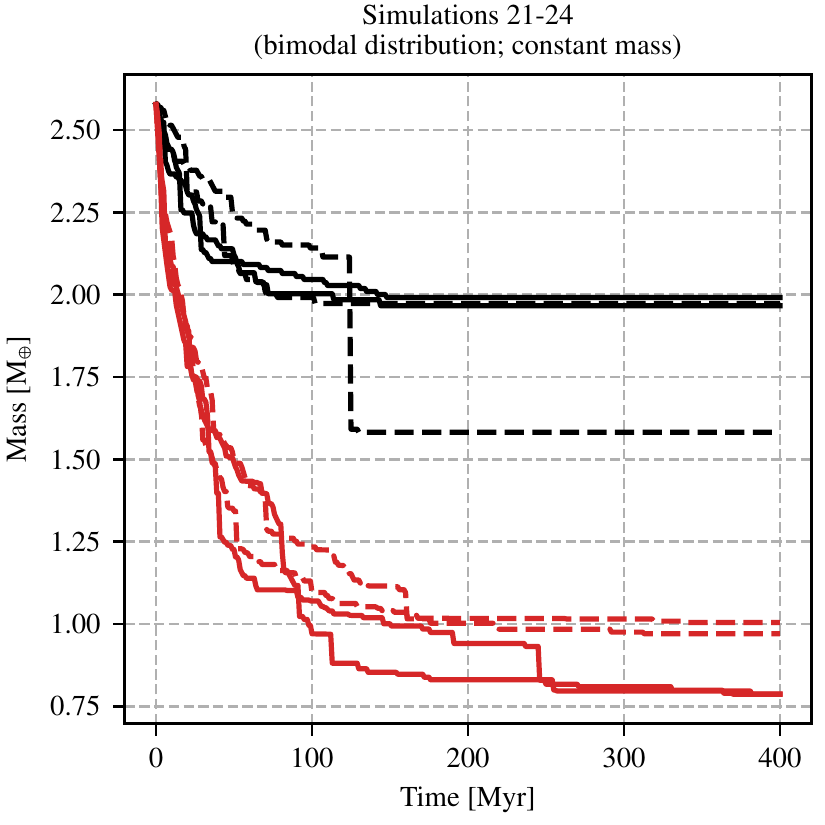}
	\includegraphics{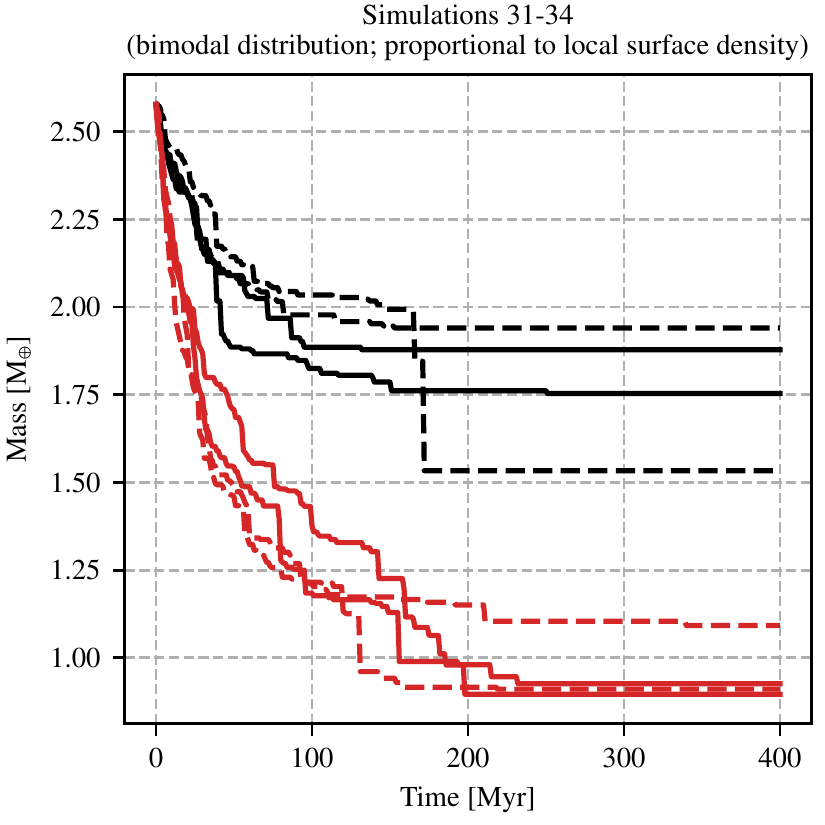}
	\caption{Total mass in the embryos as function of time for all the simulations with initial conditions from \citet{2001IcarusChambers}. Simulations are grouped by four; each group has a different initial planetesimals or embryo masses, while inside each group only the orbital elements besides the semi-major axis vary. The red lines are for the simulations with our new collision model while the black lines show the control simulations where collision are unconditionally treated as inelastic mergers.}
	\label{fig:chambers-mass}
\end{figure*}

Using our collision model, we repeat the \textit{N}-body simulations by \citet{2001IcarusChambers}. This study has the advantage of being relatively easy to perform, and has been extended and studied further by \citet{2013IcarusChambers} who adopted an idealized hit and run criterion and used scaling laws by \citet{2012ApJLeinhardt}.

Importantly, the criterion for HRC in \citet{2013IcarusChambers} is different than that used in our work. In \citet{2013IcarusChambers}, the boundary of the HRC regime is defined by a simple criterion on the impact angle, except for the high-velocity collisions. In the present work, the transition depends on impact angle and the velocity, and there is a region where HRC can occur at relatively head-on angles at intermediate velocity (see Figure~\ref{fig:classes} and the related discussion in \paperone). As a result \citet{2013IcarusChambers} over-estimates HRCs at low velocities and under-estimates HRCs at intermediate velocities.

To perform this study, we implement the present collision model in the \texttt{mercury6} \textit{N}-body code \citep{1999MNRASChambers}. The codes uses a ``radius enhancement factor'' of \num{1.2}, where a close encounter is detected if the distance between the two bodies is less than $1.2R_\mathrm{coll}$. To ensure that bodies departing a collision are not immediately identified as a close encounter, we set the relative distance factor for the resulting bodies to $d=1.25$. We perform 16 \textit{N}-body simulations similar to the ones presented in \citet{2001IcarusChambers} (that is, where collisions are treated unconditionally as mergers), and 16 others that have the same initial conditions, but using the enhanced collision model. We also made a few adaptations following \citet{2013IcarusChambers}: the time step is set to 6 days and the critical distance, closer than which a body is considered to have collided with the central star is set to \SI{0.2}{\au} due to the accuracy loss of bodies passing close to the Sun \citep{2001IcarusChambers,2013IcarusChambers}. In the simulations with the new collision model, we set a minimum mass of the remnants of \SI{e-3}{\mearth}; bodies that are below this mass are deemed to be lost.

Our goal is not to reproduce the specific attributes of the Solar System, but to understand the overall effect of a realistic collision model for the giant impact phase as compared to the similar work performed in \citet{2001IcarusChambers} and \citet{2013IcarusChambers}.

\subsection{Mass of the planets and debris production}

We show a comparison of the total mass present in the disk of embryos in Figure~\ref{fig:chambers-mass}. The simulation where collisions are treated as inelastic mergers are shown in black while the ones using the technique presented in the earlier sections are shown in red. Even with inelastic mergers, mass is still lost. The most common path is by ejections due to secular resonances with Saturn \citep{2016CeMDAHaghighipourWinter} However, the simulations using the new collision procedure lose mass more rapidly (particularly within the first 100~Myr) and in greater quantity than the control simulations. The final mass of the planets is in some cases less than $\SI{0.5}{\mearth}$, meaning that up to 80\% of the mass in the initial embryos has been lost. This indicates that unless debris reaccretion is taken into account, mass loss by collisions between embryos is more important than by the ejection of embryos or their incorporation into the Sun or gas giants. 

We observe that the initial size distribution has a non-negligible effect on the outcomes. The simulations that have a single mass distribution for the embryos (shown on the two top panels of Figure~\ref{fig:chambers-mass}) end up with lower system masses than the ones starting with a bimodal mass distribution. In the former scenario, the mass that remains in the embryos is roughly \SI{0.5}{\mearth} while in the latter the final mass ranges from 0.75 and \SI{1.25}{\mearth}. This can be explained by the types of collisions that occur for each type of simulations. 

Collisions between similar-sized bodies tend to produce more debris, whereas collisions between bodies that are more dissimilar will more likely be in the accretion regime, retaining a higher proportion of the projectile \citep{2019ApJGabriel}. Cratering events are the end-member case of this scenario, with all of the projectile being incorporated into the target. In this work, as it was similarly assumed in \citet{2013IcarusChambers}, we neglect the reaccretion of debris.

\subsection{Types of collisions}

\begin{table}
    \centering
    \caption{Breakdown of collision types obtained during the \textit{N}-body simulations.}
    \begin{tabular}{lcccc}
        \hline
        & \multicolumn{4}{c}{Simulations} \\
        Type & 01-04 & 11-14 & 21-24 & 31-34 \\
        \hline
        Merger & 472 & 446 & 427 & 361 \\
        \hline
        Accretion & 231 & 233 & 170 & 155 \\
        Erosion & 156 & 124 & 146 & 132 \\
        Hit-and-run & 479 & 504 & 253 & 213 \\
    \end{tabular}
    \tablecomments{The values in the ``Merger'' line are the collisions in the simulations where collisions are unconditionally as merges. The other three lines are for the collisions obtained in the simulations with the new model described in this work. The classification of the latter is made using the method outlined in Section~\ref{sec:classifier}.}
    \label{tab:coll-types}
\end{table}

We provide a comparison of the number of collisions and their types in Table~\ref{tab:coll-types}. The first line contains the total number of collisions that are obtained in the simulations where all collisions are treated as mergers while the other are for the simulations with the collision model introduced earlier in this work.

It can be seen that with the new collision model, the number of collisions is greatly increased. In the cases where all the initial bodies are of comparable mass (simulations 01 to 14), the number is almost doubled. In those simulations, the predominant regime is HRCs. The fraction of HRCs is 55\%, which is slightly higher than the 49\% obtained by \citet{2010ApJKokubo} for a similar situation (only embryos). The simulations with a bimodal mass distribution (number 21 to 34) show both a reduced number of collisions and a reduced fraction of HRCs. For these simulations, we find 44\% of the collisions to be HRC, consistent with the 42\% found by \citet{2013IcarusChambers} for embryo-embryo collisions.

The increased number of HRCs in the simulations with similar-mass bodies can be explained by the geometry. With equal size bodies, about 3/4 of the collisions are grazing using the criterion of \citet{2010ChEGAsphaug}, while when $r_\mathrm{P}\simeq r_\mathrm{T}/2$ (a sufficient approximation for $\gamma=0.1$), the fraction of grazing collisions reduces to about the half. The increased number of collisions also has implications for the production of debris. As nearly every collision produces some debris, the more numerous the collisions, the lower the mass that remains in the embryos.

\subsection{Radial mixing}

\begin{figure*}
	\centering
	\includegraphics{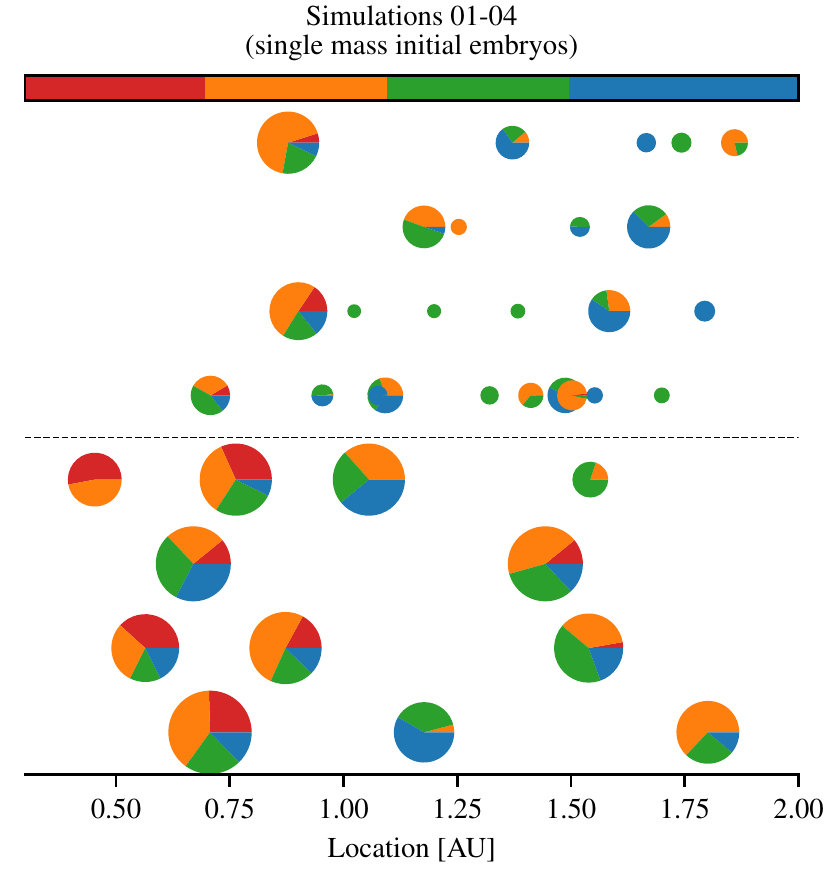}
	\includegraphics{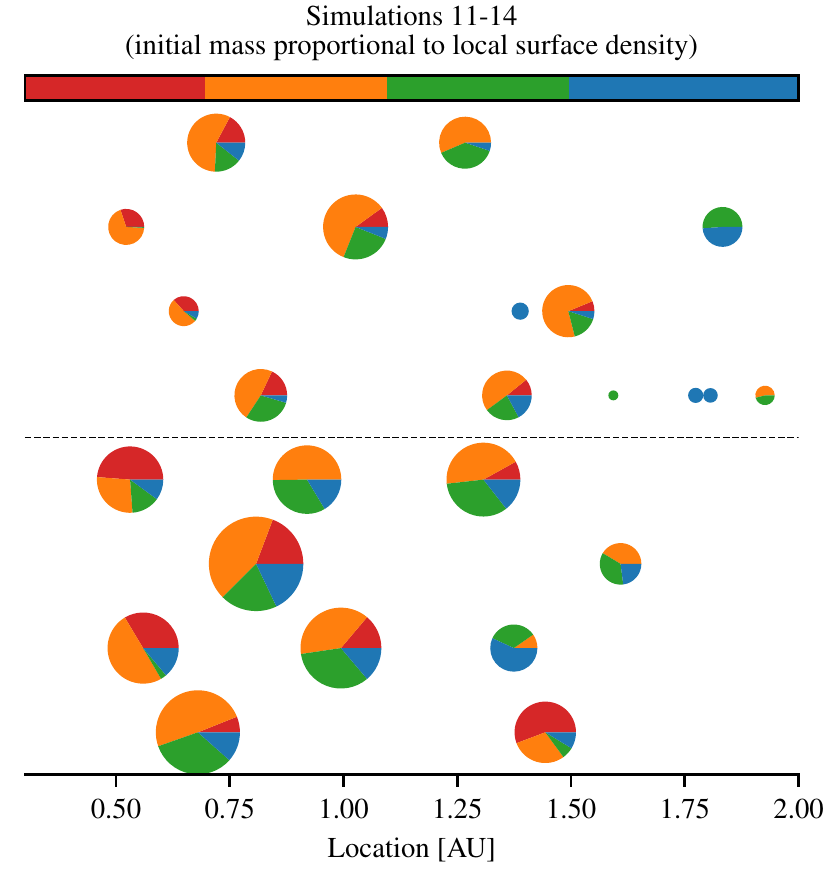}
	\includegraphics{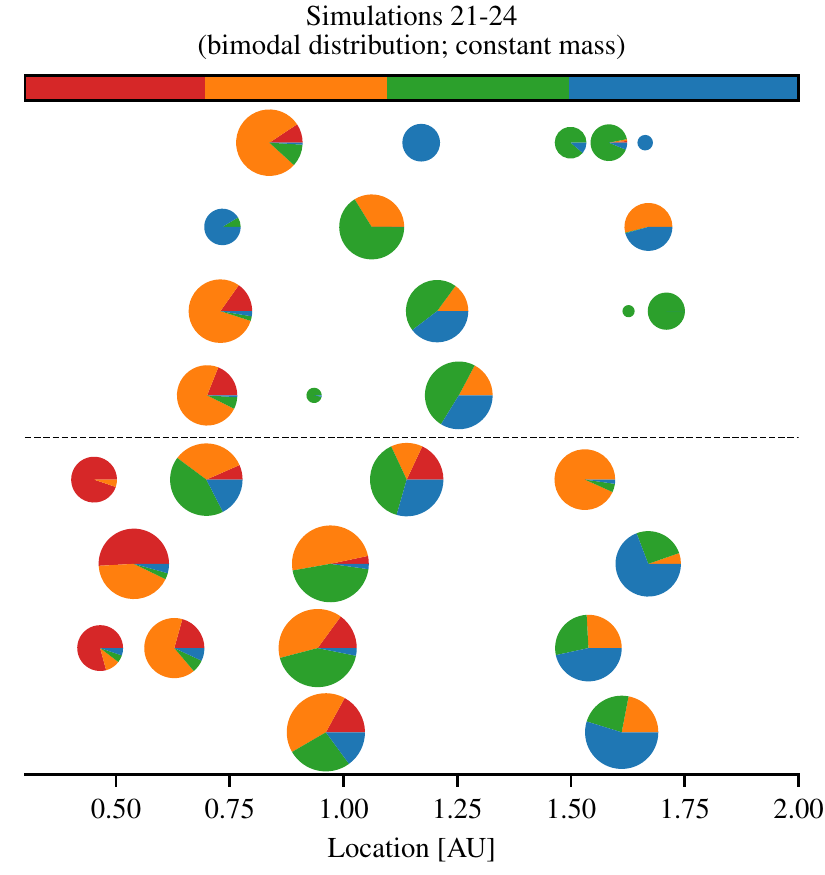}
	\includegraphics{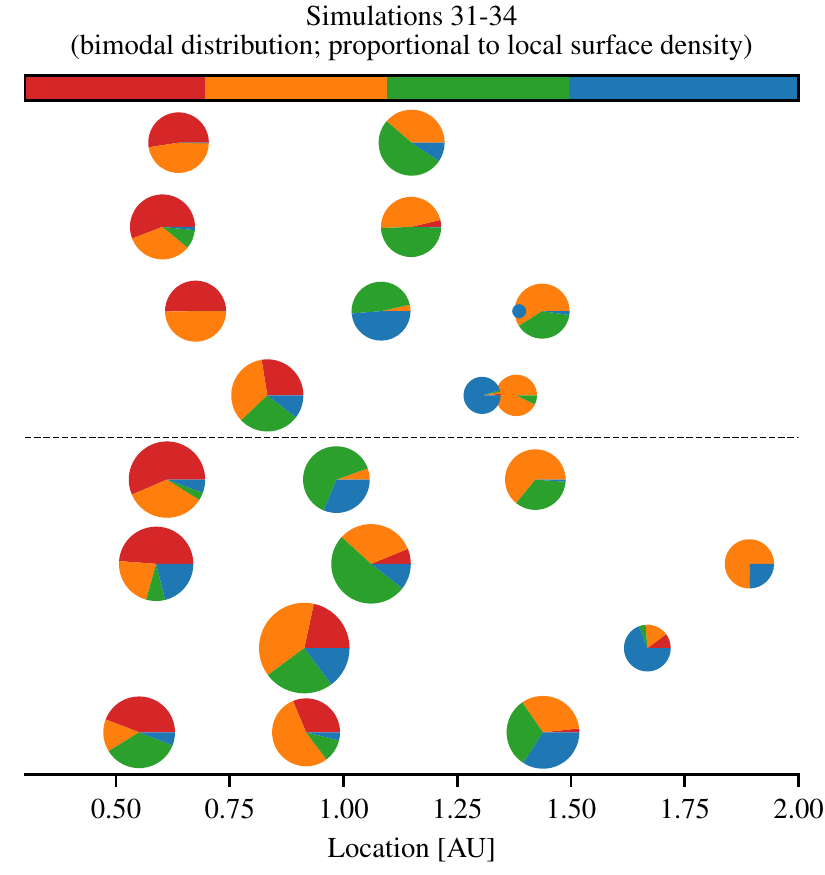}
	\caption{End state of the 32 \textit{N}-body simulations that are part of this work. Each system is shown on one line, with the resulting planets placed at their semi-major axis and whose radius is proportional to their physical radius assuming a constant density. The resulting bodies are colored by fraction of mantle material coming from four different regions (whose boundaries are show on the top). In each panel, the top four systems were computed using the realistic collision model whereas the bottom four were computed assuming collisions are mergers. Smaller planets are placed in front of larger ones so there are no hidden objects.}
	\label{fig:chambers-compo}
\end{figure*}

To assess the location in the disk of the material that forms the final bodies, we adopt a simple procedure similar to that of \citet{2001IcarusChambers} and \citet{2013IcarusChambers}. The initial bodies are divided into four compositional zones: interior to \SI{0.7}{\au} (red), from \num{0.7}--\SI{1.1}{\au} (orange), from \num{1.1}--\SI{1.5}{\au} (green), and beyond \SI{1.5}{\au} (blue). To determine the composition change during a collision, in the case of erosion, the mantle is eroded first, and the core is eroded only if no mantle is left, as in \citet{2010ApJMarcus}. When accretion occurs, the mass difference is taken from the second body (the projectile) starting with its mantle. HRCs are treated in the same way, with erosion from the other body serving as the material source in case a body has a positive accretion efficiency.

Figure~\ref{fig:chambers-compo} shows the final states of the accreted planetary systems, along with the described material origin of each planet (red-orange-green-blue), comparing outcomes for perfect accretion (the bottom four systems in each of the four panels) with our more realistic treatment of giant impacts (the top four systems in each panel). The final systems in each panel are the result of the mass evolution given by the black and red curves in each panel in Figure~\ref{fig:chambers-mass}. 

First we note the differences in the final systems when the starting bodies have a similar mass (the top two panels) compared with starting bodies following a bimodal mass distribution (bottom two panels). When the starting bodies are similar sized, early collisions include a preponderance of HRCs whose smaller remnants survive until the end of the integration. When starting bodies are bimodal, the systems more closely resemble the ones obtained assuming perfect merger in each collision. This is due to the fact that HRC is more common when bodies are similar-sized due to the greater fraction of material not intersecting in a given collision \citep{2012ApJLeinhardt,2016IcarusMovshovitz,2019ApJGabriel}.

The neglect of debris reaccretion in our simulations means that these initial results should be taken cautiously. About 1--$\SI{2}{\mearth}$ of material is not conserved in our simulations, when in reality a large fraction of this may be reaccreted over time onto the growing embryos \citep{2012MNRASJacksonWyatt}. This means that the planets obtained in the simulations with our collision model are most likely too small.

Systems with smaller masses have generally longer formation times \citep{2006ApJKokubo}. Hence, we expect that the systems where the are still numerous bodies haved not reached their final state at the time we end our integrations (\SI{400}{\myr}). Figure~\ref{fig:chambers-mass} also suggest that the frequency of collisions decreases at later times, as indicated by the flattening of the red curves in. As the collision rate slows down towards the later times (See Figure~\ref{fig:chambers-mass}), it could take hundreds of millions of years or longer for most of the smaller bodies to be accreted or removed.

\subsection{Material reprocessing by HRC}

\begin{figure}
    \includegraphics{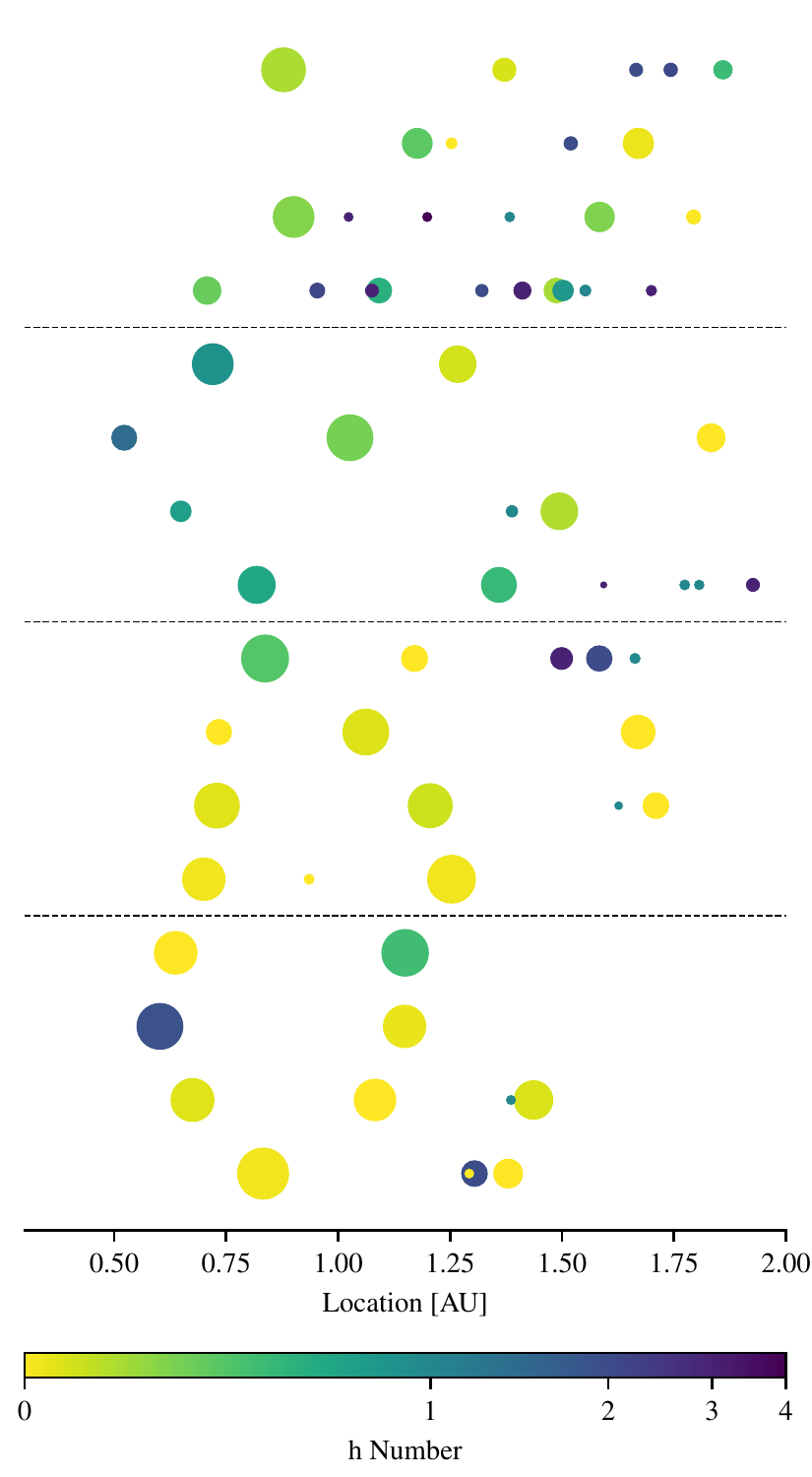}
    \caption{End state of the 16 $N$-body simulations featuring the surrogate collision model, now shown as one system per line (simulations 01 to 34 from top to bottom). As before, the size of each planet is shown proportional to its physical size assuming a constant density. The first 8 begin with comparable-sized embryos, while the last 8 begin with a bimodal mass distribution of embryos. Here each final planet is colored by the number of hit and run collisions it has experienced, where the h-number is mass-averaged during each accretion event as described in the text.}
    \label{fig:chambers-hnum}
\end{figure}

In perfect merger simulations, there is a comparative lack of less massive bodies. Although some of the smaller remnant planets in our simulations are unaccreted original embryos, most of them are the ``runners'' from HRCs (see Figure~\ref{fig:chambers-hnum}), so some further discussion is necessary. 

In Figure~\ref{fig:chambers-hnum} we quantify the cumulative occurrence of HRCs during each planet's formation, for all the simulations featuring the realistic collision model. Each line represents the final outcome of one simulation. The results are divided in four groups, each containing simulations that were initialized using a different method to determine the initial masses of the embryos, as per Figure~\ref{fig:chambers-compo}, where first 8 sets of runs begin with initial embryos of the same or similar mass, while the last 8 sets of runs start with a bimodal mass distribution. In each case the initial embryos all start with ``h-numbers'' of zero. When a HRC occurs, the surviving projectile (runner) has its $h$ increased by 1. When there is a merger, the resulting $h$ is the mass average of $h$ in each progenitor. 

In our simulations, we obtain values for $h$ in the range of 0 to 4; the color scale is linear from 0 to 1 and logarithmic above. We also find that smaller bodies tend to have larger h-numbers than bigger bodies, indicating the dynamical survival of runners. There are a couple of simulations where a few smaller bodies remain until the end of our integration (\SI{400}{\myr}) after having survived up to 5 consecutive HRCs. It is possible that several from these bodies will be accreted further on, as they remain close to other more massive bodies.

\subsection{Composition}

Figure~\ref{fig:chambers-compo} shows that the smaller mass planets in our simulations originate primarily from a single region represented by a single color. Although our prescription for material transfer is relatively basic for an HRC, as runners are assumed to be made of projectile material, it is largely consistent with the SPH simulations of HRCs that produce only a limited amount of equilibration between the colliding bodies. The limited amount of mixing during HRCs also gives the possibility of transporting projectile runners for large distances without equilibration, as was already pointed out in \citetalias{2019ApJEmsenhuberA}, where it was found that the final destination of a runner does not depend on its pre-impact trajectory.

For most of the simulations, we also note that the resulting planets are composed of only a small fraction of material coming from the inner region (shown in red in Figure~\ref{fig:chambers-compo}). Unlike \citet{2013IcarusChambers}, we find that material coming from the inner part of the original system has a low survival rate. 

To understand this difference, we note that when collisions are treated unconditionally as mergers, the only possibility for mass loss is through ejection from the solar system or collision with the sun. Selective ejection of outer bodies (blue) is anticipated because they have a lower escape velocity from the solar system and can be strongly perturbed gravitationally by gas giants. On this basis we would expect less mass loss from the inner solar system. But when debris production is included---inefficient accretion of our surrogate model---there is a bias towards mass loss from the inner region due to the greater absolute impact velocities. Impact velocity is generally greater for closer-in collisions in proportion to their higher Keplerian velocity $v_\mathrm{kep}=\sqrt{GM_\star/a}$. 

We caution that some of these conclusions may be an artifact of our simple treatment of debris, which we treat as being lost, when in fact the debris that is closest to the sun would be swept up the fastest by the remaining planets since the dynamical time scale is shorter. Careful treatment of the debris in our \textit{N}-body scheme is an important matter for future work. In addition to influencing the cumulative mass, the production of debris would lead to increased dynamical friction which would reduce the eccentricities and inclinations of the embryos \citep{2006IcarusRaymond,2006IcarusOBrien,2019arXivKobayashi}, and hence their relative velocities, increasing the accretion efficiency but reducing the interactions across feeding zones. 

% =======================
\section{Conclusions}
% =======================

We use a surrogate model for giant impacts (\citet{2019ApJCambioni} and this work), trained on hundreds of SPH simulations from \citet{2011PhDReufer} in the regime applicable to ``late stage'' terrestrial collisions, to explore aspects of planet formation that could not be studied before: the loss of mass due to imperfect accretion, and the evolution of survivors of HRCs. 

The surrogate model we have used in this study is available for public use; see Section~\ref{sec:collresolve}. Updates will be published as more data sets become available. We plan to extend the training data to the regime of less than a few thousands kilometers in diameter, when friction matters \citep{2018AASugiura,2015PSSJutzi}. Pre-collisional rotation represents a significantly larger expansion of the parameter space \citep{2008IcarusCanup}, introducing several additional free parameters per collision; this could require 100 times as many simulations if sampled adequately, which is currently prohibitive. For now, pre-impact rotation is perhaps better suited to physical scaling including aspects related to composition and differentiation \citep{2019ApJGabriel}, that could be benchmarked in some cases and applied to the output of non-initial-rotation surrogate models. And except in the simplest sense (e.g. Figure~\ref{fig:chambers-compo} we have not analyzed how composition changes due to mantle stripping by each HRC \citep{2014NatGeoAsphaug}.

In the spirit of \citet{2010ApJKokubo}, \citet{2013IcarusChambers} and other studies we have evolved several archetypal scenarios of solar system terrestrial planet formation to understand the process of embryo accretion and its consequences. In this first basic application of our surrogate model, we find that the inclusion of realistic collisions changes the dynamics of planetary growth in important ways. 
First, planet formation takes significantly longer to complete when accounting for realistic collisions. The phase of excited dynamics and evolving compositions continues long after the end of our simulations, and has significant implications for how accretion ends, e.g. the late veneer \citep{2015GMSMorbidelliWood} and the idea of the late heavy bombardment \citep{2019GeoSciHartmann}, that we have not yet studied.
Second, realistic collisions yield a much greater diversity of finished planets, in terms of their sizes and compositions, because there are surviving runners and their byproducts, subject to mantle stripping. When perfect merger is assumed, it is possible to directly merge planets from distant feeding zones; these mergers would actually be HRCs with only limited material capture.
Third, we find that runners from HRCs are common in the final population of bodies, and that multiple-HRC survivors are expected to exist, as proposed by \citet{2014NatGeoAsphaug} and \citet{2018ApJChau} for the origin of Mercury. 

Finally, we have not included any specific evolution of the giant planets in our model. This is reasonable, because the terrestrial planets most likely formed after possible instabilities of the giant planets. \citet{2018IcarusMorbidelli} found that the abundances of highly siderophile elements in the lunar mantle can be explained by the accretion tail from the formation of the terrestrial planets. The presence of binary Jupiter trojans has also been interpreted as an evidence for a very early migration \citep{2018NatAsNesvorny}. Both these studies favor an early instability, if it ever took place.

\acknowledgements

A.E., S.C., E.A. and S.R.S. acknowledge support from NASA under grant 80NSSC19K0817.
We thank the anonymous reviewer for the precious comments and and edits that improved this manuscript.
An allocation of computer time from UA Research Computing High Performance Computing (HPC) is gratefully acknowledged.

\software{Mercury \citep{1999MNRASChambers}, matplotlib \citep{2007CSEHunter}}

\bibliographystyle{aasjournal}
\bibliography{bib}

\end{document}